\documentstyle[epsf,12pt]{article}

\makeatletter

 \@addtoreset{equation}{section}
\makeatother

\begin{document}
\topmargin 0pt
\oddsidemargin 5mm

\newcommand {\beq}{\begin{eqnarray}}
\newcommand {\eeq}{\end{eqnarray}}
\newcommand {\non}{\nonumber\\}
\newcommand {\eq}[1]{\label {eq.#1}}
\newcommand {\defeq}{\stackrel{\rm def}{=}}
\newcommand {\gto}{\stackrel{g}{\to}}
\newcommand {\hto}{\stackrel{h}{\to}}

\newcommand {\1}[1]{\frac{1}{#1}}
\newcommand {\2}[1]{\frac{i}{#1}}

\newcommand {\th}{\theta}
\newcommand {\thb}{\bar{\theta}}
\newcommand {\ps}{\psi}
\newcommand {\psb}{\bar{\psi}}
\newcommand {\ph}{\varphi}
\newcommand {\phs}[1]{\varphi^{*#1}}
\newcommand {\sig}{\sigma}
\newcommand {\sigb}{\bar{\sigma}}
\newcommand {\Ph}{\Phi}
\newcommand {\Phd}{\Phi^{\dagger}}
\newcommand {\Sig}{\Sigma}
\newcommand {\Phm}{{\mit\Phi}}
\newcommand {\eps}{\varepsilon}
\newcommand {\del}{\partial}
\newcommand {\dagg}{^{\dagger}}
\newcommand {\pri}{^{\prime}}
\newcommand {\prip}{^{\prime\prime}}
\newcommand {\pripp}{^{\prime\prime\prime}}
\newcommand {\prippp}{^{\prime\prime\prime\prime}}
\newcommand {\delb}{\bar{\partial}}
\newcommand {\zb}{\bar{z}}
\newcommand {\mub}{\bar{\mu}}
\newcommand {\nub}{\bar{\nu}}
\newcommand {\lam}{\lambda}
\newcommand {\lamb}{\bar{\lambda}}
\newcommand {\kap}{\kappa}
\newcommand {\kapb}{\bar{\kappa}}
\newcommand {\xib}{\bar{\xi}}
\newcommand {\Ga}{\Gamma}
\newcommand {\rhob}{\bar{\rho}}
\newcommand {\etab}{\bar{\eta}}
\newcommand {\tht}{\tilde{\th}}
\newcommand {\zbasis}[1]{\del/\del z^{#1}}
\newcommand {\zbbasis}[1]{\del/\del \bar{z}^{#1}}

\newcommand {\vecv}{\vec{v}^{\, \prime}}
\newcommand {\vecvd}{\vec{v}^{\, \prime \dagger}}
\newcommand {\vecvs}{\vec{v}^{\, \prime *}}

\newcommand {\alpht}{\tilde{\alpha}}
\newcommand {\xipd}{\xi^{\prime\dagger}}
\newcommand {\pris}{^{\prime *}}
\newcommand {\prid}{^{\prime \dagger}}
\newcommand {\Jto}{\stackrel{J}{\to}}
\newcommand {\vprid}{v^{\prime 2}}
\newcommand {\vpriq}{v^{\prime 4}}
\newcommand {\vt}{\tilde{v}}
\newcommand {\vecvt}{\vec{\tilde{v}}}
\newcommand {\vecpht}{\vec{\tilde{\phi}}}
\newcommand {\pht}{\tilde{\phi}}

\newcommand {\goto}{\stackrel{g_0}{\to}}
\newcommand {\tr}{{\rm tr}\,}

\newcommand{\vs}[1]{\vspace{#1 mm}}
\newcommand{\hs}[1]{\hspace{#1 mm}}

\setcounter{page}{0}

\begin{titlepage}

\begin{flushright}
KEK-TH-572\\
hep-th/9805038\\
May 1998
\end{flushright}
\bigskip

\begin{center}
{\LARGE\bf
Moduli Space of Global Symmetry in\\
$N=1$ Supersymmetric Theories and\\
\bigskip\medskip
the Quasi-Nambu-Goldstone Bosons
}
\vs{10}

\bigskip
{\renewcommand{\thefootnote}{\fnsymbol{footnote}}
{\large\bf Muneto Nitta\footnote{
muneto.nitta@kek.jp, nitta@phys.wani.osaka-u.ac.jp.}
}}

\setcounter{footnote}{0}
\bigskip

{\small \it
Department of Physics,
Graduate School of Science, Osaka University,\\
Toyonaka, Osaka 560-0043, Japan  and\\
\medskip
Theory Division,
Institute of Particle and Nuclear Studies, KEK,\\
Tsukuba, Ibaraki 305-0801, Japan
}
\end{center}
\bigskip

\begin{abstract}
We derive the moduli space for the global symmetry 
in $N=1$ supersymmetric theories.
We show, at the generic points, 
that it coincides with the space of quasi-Nambu-Goldstone (QNG) bosons, 
which appear besides the ordinary Nambu-Goldstone (NG) bosons 
when the global symmetry $G$ breaks down spontaneously  
to its subgroup $H$ with preserving $N=1$ supersymmetry.
At the singular points, 
most of the NG bosons change to the QNG bosons 
and the unbroken global symmetry is enhanced. 
The $G$-orbits parametrized by the NG bosons 
are the fibre at the moduli space and 
the singular points correspond to the point 
where the $H$-orbit (in the $G$-orbit) shrinks.
We also show that the low-energy effective Lagrangian  
is an arbitrary function of the moduli parameters.
\end{abstract}

\end{titlepage}

\newpage
\section{Introduction}

In this paper, 
we investigate the relation between two elements: 
one is the moduli space of $global$ symmetry in 
the supersymmetric (gauge) theories; 
the other is the low-energy effective Lagrangian 
described by the supersymmetric nonlinear sigma model 
with the K\"{a}hler target manifold parametrized by 
the chiral Nambu-Goldstone (NG) superfields. 
The moduli space of gauge symmetry is well understood 
in terms of the K\"{a}hler quotient space~\cite{HKLR} or 
the the algebraic variety~\cite{LT,DM}. 
The zeros of the scalar potential is made by 
the F-flat condition and the D-flat condition. 
Since the gauge symmetry $G$ is enhanced to 
its complexfication $G^{\bf C}$, 
it is easy to deal with.
However, in the case of global symmetry $G$, 
although the F-term symmetry is enhanced to 
its complexification by 
the analyticity of the superpotential, 
the D-term symmetry is $not$ enhanced to 
its complexification, 
since it includes both chiral and anti-chiral superfields. 
Therefore, the moduli space of global symmetry 
is obtained by the set of F-flat points 
divided by the symmetry $G$, but not its complexification, 
since there is no D-flat condition.
Since it is not well understood, 
we investigate it in this paper.
The F-term zeros is just the $G^{\bf C}$-orbit of the vacuum. 
In the case of gauge symmetry, 
since the D-flat points constitute 
the one $G$-orbit in the F-flat points, 
the moduli space of gauge symmetry is parametrized by 
$G^{\bf C}$-invariant polynomials. 
In the case of global symmetry, 
the $G$-invariant, but not $G^{\bf C}$-invariant, polynomials 
parametrize the moduli space. 

On the other hand, 
the F-term zeros is well known in the context of 
the low-energy effective Lagrangian. 
When global symmetry breaks down spontaneously to its subgroup, 
there appear quasi Nambu-Goldstone (QNG) bosons besides 
the ordinary Nambu-Goldstone (NG) bosons as massless bosons. 
They constitute massless NG chiral superfields 
with QNG fermions (their fermion partners).
After integrating out the massive mode, 
the low-energy effective Lagrangian is obtained as 
a supersymmetric nonlinear sigma model with 
the K\"{a}hler target 
manifold~\cite{Zu,BKMU,Le,BL,Sh,KS,LRM,IKK,BE,HNOO,HN}. 
(For a review see \cite{Ku,BKY,WB}.)
Since the target space is the F-term zeros, 
it is a K\"{a}helr coset manifold, 
where $G^{\bf C}$ acts transitively, but not isometrically, and 
$G$ acts isometrically, but not transitively. 
The general K\"{a}hler potential with $G$ symmetry has been 
obtained by Bando, Kuramoto, Maskawa and Uehara (BKMU) 
in Ref.~\cite{BKMU}.
It has not been known up to recently 
that low energy theorems exist~\cite{HNOO,HN}, 
since the K\"{a}hler potential includes 
an arbitrary function of $G$-invariants.
However, since it has not been known  
how many variables are included in it, in general, 
except for few examples~\cite{KS}, 
we determine the number of variables in 
the arbitrary function of the effective K\"{a}hler potential.
The number of QNG bosons changes, 
even if we consider one theory, 
since there exist 
supersymmetric vacuum alignment~\cite{KS2,KS}.
Although it is known that 
there must exist at least one QNG boson~\cite{Le,BL,KS}, 
the minimum number of QNG bosons has not been known, 
except for few examples~\cite{KS}.
We determine the range of the number of QNG bosons 
and show that the minimum number of QNG bosons coincides with 
the number of variables in the arbitrary function and 
the dimension of the moduli space of global symmetry.

We investigate the moduli space of global symmetry in two ways: 
one involves algebraic-geometrical methods, 
such as the algebraic variety; 
the other involves the differential-geometrical or 
group-theoretical methods, 
such as the K\"{a}hler coset manifolds. 
The former is used mainly to calculate 
the dimension of the moduli space; 
the latter is used to 
investigate more complicated structures of the moduli space.

\bigskip
This paper is organized as follows. 
In the next section, 
we discuss the general aspects of moduli space 
in both the gauge and global symmetry. 
We found, in the case of global symmetry, 
that the moduli space is the quotient space, 
the set of F-flat points divided by the symmetry.
Since the set of F-flat points is 
the $G^{\bf C}$-orbit, known as the K\"{a}hler coset manifold, 
we review only its basic aspects. 

In Sec.~3, we investigate 
the detailed structure of moduli space 
in the algebraic-geometrical and 
the differential-geometrical ways, 
and show the relation to 
the low-energy effective Lagrangian 
describing the behavior of the NG and QNG bosons. 
In Sec.~3.1, we show that 
the moduli space of global symmetry 
is parametrized by moduli parameters, 
which are $G$-invariants, but not $G^{\bf C}$-invariants.  
We decompose the moduli space into several regions 
with isomorphic unbroken symmetries. 
Sec.~3.2 is devoted to an investigation of 
the low-energy effective Lagrangian.
We show that 
it can be written as an arbitrary function of the moduli parameters, 
and that it is equivalent to the known K\"{a}hler potential~\cite{BKMU,Le} 
by identifying the moduli parameters constrained by the F-term 
and the representative of K\"{a}hler coset manifold. 
(Thus the number of variables is the dimension of moduli space.)  

In Sec.~3.3, we investigate 
the structure of the K\"{a}hler coset manifold in detail. 
The main results of this paper are obtained in this subsection. 
We generalize the observation by Kotcheff and Shore~\cite{Sh,KS} 
that the non-compact directions belonging to 
the same $H$-irreducible sector are not independent. 
It is shown that the number of $G$-invariants agrees with 
the number of $H$-irreducible sectors of 
the mixed-type complex broken generators, 
but in general it does not coincide with 
the dimension of moduli space. 
We also show that the dimension of each region of moduli space 
is the number of $H$-singlet sectors. 
We prove that, in a generic region, 
all $H$-irreducible sectors become singlet, 
and that their number is just the dimension of moduli space.
We also show that the singular points in moduli space 
correspond to the points where the $H$-orbit shrinks in 
the target manifold. 
Sec.~3.4 is devoted to a calculation of 
the dimension of moduli space and the number of QNG bosons. 

In Sec.~4, 
we give some examples and 
demonstrate the theorems obtained in Sec.~3. 
We investigate $O(N)$ and $SU(N)$ with 
fundamental (and anti-fundamental) matters 
and $SU(N)\;(N=2,3)$ with adjoint matters. 
In Sec.~5, 
we make some comments concerning the gauging of global symmetry. 
Sec.~6 is devoted to conclusions.

\section{Moduli space}

In this paper, we assume that 
the vacua are transformed by some symmetry $G$. 
Suppose that there is a gauge and global symmetry, 
$G = G_{\rm gauge} \times  G_{\rm global}$.
Although it does not have to be direct product, 
for simplicity we suppose it here.  
The fundamental field $\vec{\phi}$ is 
in some representation space, $V = {\bf C}^N$ of $G$.
The scalar potential is~\cite{WB}
\beq
 {\cal V}(\phi,\phi^*) 
 = \1{2} (\vec{\phi}\dagg T_A \vec{\phi})^2 + |W\pri|^2 ,
\eeq
where $T_A$ is the generator of the gauge group. 
The first term comes from the $D$-term of the gauge symmetry 
and the second term from the F-term.
The D-flat condition is necessary and sufficient condition that 
the length of the vector $\vec{\phi}$, $|\vec{\phi}|$, 
is minimum~\cite{D-flat,GS}.
Since the moduli parameters are the freedom 
that remains after bringing the fields $\vec{\phi}$ 
to some constant configuration
by using all of the gauge and the $global$ symmetry, 
the moduli space is written as~\footnote{
There is another definition of the moduli space.
Sometimes the moduli space is defined 
without the global symmetry 
since the quotient by the global symmetry is not easy.
In such a definition, the NG bosons are included in 
the moduli space.}   
\beq
 {\cal M} 
&=& \{\vec{\phi}\in {\bf C}^N \,|\, {\cal V}(\phi,\phi^*) = 0 \}/G \non
&=& \{\vec{\phi} \,|\, W\pri = 0 \,,\, 
        { |\vec{\phi}|^2 = {\rm min.}} \}/G \non
&=& \{\vec{\phi} \,|\, W\pri = 0 \}
     /{ G_{\rm gauge}^{\bf C}}\times G_{\rm global} \;, 
\eeq
where we have used the fact (see \cite{WB})
\beq
&& \{\, \cdot \,|\, \mbox{D-flat cond.} \}/G_{\rm gauge} 
    = \{\, \cdot \,\}/G_{\rm gauge}^{\bf C} \;,
\eeq
which is a result of the Higgs mechanism.
The D-flat condition corresponds to 
$G_{\rm gauge}^{\bf C}/G_{\rm gauge}$, 
which is also equivalent to 
taking the Wess-Zumino gauge~\cite{WB,LT}.

Here, we consider the two extreme cases.
First of all, consider the case 
without global symmetry, $G=G_{\rm gauge}$. 
In this case, the moduli space is 
\beq
 {\cal M} 
 = \{\vec{\phi}\,|\, W\pri = 0 \}/G^{\bf C} .\label{KQ}
\eeq
This is known as 
the K\"{a}hler quotient~\cite{HKLR}, 
and can be understood as the algebraic variety~\cite{LT,DM}.
It is parametrized by 
the $G^{\bf C}$-invariant polynomials in $V$, 
which are elements of 
the ring of the invariant polynomials, $A^{G^{\bf C}}[V]$.
However for the case where the zeros of the potential can be 
transformed by the symmetries, 
it becomes trivial, ${\cal M} \simeq \{1\}$, 
since there is no global symmetry.~\footnote{
In the another definition of the moduli space, 
Eq.~(\ref{KQ}) is obtained as the moduli space 
when there is also global symmetry or 
not supposing that the vacua should be transformed by the symmetry.
See Ref.~\cite{LT,DM}.}

We thus consider another extreme case, 
where there is no gauge symmetry, $G=G_{\rm global}$.
In this paper we investigate this case in detail.
(If there is a gauge symmetry, 
we consider them as global symmetry for a while, 
and then gauge them after finding the moduli space.
This point of view is discussed in Sec.~5.) 
In this case, the moduli space is
\beq
 {\cal M} = \{\vec{\phi}\,|\, W\pri=0 \}/G .
\eeq
The zeros of the F-term potential\footnote{
Since we are using the effective Lagrangian approach, 
we discuss the superpotential $W$ 
with quantum corrections, if any.}, 
$M = \{\vec{\phi}\,|\, W\pri=0 \}$, is obtained 
when all $G^{\bf C}$-invariant polynomials are fixed (see Fig.~1).
\begin{figure}
 \epsfxsize=6cm
 \centerline{\epsfbox{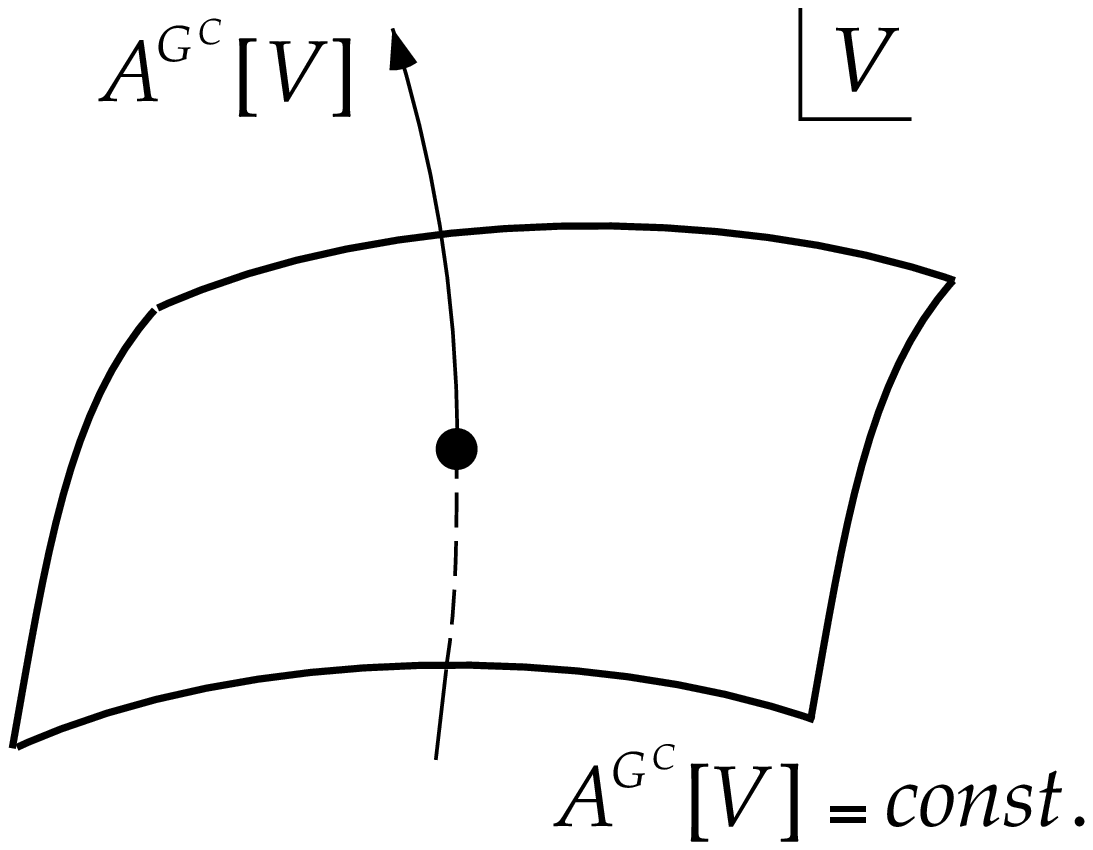}}
 \centerline{\mbox{Figure 1 : {\bf zeros of F-term potential}}}
  \begin{center}
  \begin{footnotesize}
The target manifold $M$ is defined as the zeros of F-term potential.
  \end{footnotesize}
  \end{center}
\end{figure}
Since we consider the case when any vacuum 
is transformed by some (complexified) symmetry, 
$G^{\bf C}$ acts on $M$ transitively, 
so that $M$ can be written in complex coset space as
\beq
 M \simeq G^{\bf C}/\hat H ,  \label{coset}
\eeq
where $\hat H$ is the complex isotropy group 
at the vacuum, $\vec{v} = <\vec{\phi}>$, namely 
\beq
 \hat H_v = \{g \in G^{\bf C}|g \cdot \vec{v} = \vec{v} \}.
\eeq
(We have omitted $v$ in Eq.~(\ref{coset}), 
since $G^{\bf C}$ is transitive 
and each $\hat H_v$ is isomorphic.)
Here, $\hat H$ includes $H^{\bf C}$, 
$\hat H \supset H^{\bf C}$, 
but need not agree with it~\cite{BKMU,GS}. 
At special points in $M$, 
it can be decomposed as
\beq
 \hat{\cal H} = \cal H^{\bf C} \oplus \cal B \;, \label{Hhat-dec.}
\eeq
where $\cal B$ is the nilpotent Lie algebra, 
which is written in the non-Hermitian 
step generators, namely, 
the lower-half triangle matrices in 
the suitable basis~\cite{BKMU}(see Sec.~4.2).
We always omit the subscript $v$ on $H$ 
at such points.
When ${\cal B}$ is absent, $\hat H$ is called reductive.
${\cal B}$ is determined completely 
by the representations to which the vacuum vectors belong. 
$M$ is the target space of the sigma model 
parametrized by the NG and 
QNG bosons~\cite{BKMU,Le,BL,Sh,KS,HNOO,HN}.
The moduli space is written as 
\beq 
 {\cal M}  = (G^{\bf C}/\hat H) /G \;.
\eeq
This is a quotient space, but not a K\"{a}hler quotient space 
(for a review of quotient space, see Ref.~\cite{HKLR}).
Naively, the moduli space is parametrized by the QNG bosons, 
since the NG bosons correspond 
to compact directions generated by 
the compact isometry group $G$.

In this paper we consider the generic $G^{\bf C}$-orbit 
which has the maximal dimension. 
(Generalization to the singular $G^{\bf C}$-orbits, 
which has fewer dimensions, is straightforward.)
The number $N_{\Phi}$ of the massless NG chiral multiplets 
parametrizing $M$ is (see Fig.~1, see also for example \cite{DM})~\footnote{
We use symbols `dim' for a real dimension and `${\rm dim}_{\bf C}$' 
for a complex dimension.} 
\beq
 N_{\Phi} = \dim_{\bf C} (G^{\bf C}/\hat H) 
 = \dim_{\bf C} V - N(G^{\bf C}) \;,\label{NPhi} 
\eeq
where $N(G^{\bf C})$ is the number of 
$G^{\bf C}$-invariant polynomials, namely, 
\beq
 N(G^{\bf C}) \defeq \dim_{\bf C}(A^{G^{\bf C}} [V]) \;. 
\eeq

There are two types of the chiral NG multiplets~\cite{BKMU,Le}.
One is called the $pure$ $type$ (or non-doubled type),
including two NG bosons in the scalar component.
Another is called the $mixed$ $type$ (or doubled type), 
including the QNG boson besides the ordinary NG boson.
The mixed types correspond to Hermitian generators, 
whereas the pure types correspond to non-Hermitian generators.
(We can always obtain ordinary Hermitian generators 
by suitable complex linear combinations of 
the pure-type broken generators and 
the Borel-type unbroken generators ${\cal B}$ 
in the complex unbroken algebra $\hat{\cal H}$.)
$N_{\Phi}$ can be written as
\beq
 N_{\Phi} = N_{\rm M} + N_{\rm P} \;,
\eeq
where $N_{\rm M}$ and $N_{\rm P}$ are 
the number of the mixed-type and the pure-type multiplets.

In general, $N_{\rm M}$ and $N_{\rm P}$ can change 
at various points in target space $M$, 
with the total numbers being conserved.
This is because there is a vacuum alignment~\cite{KS2,KS}.
\begin{figure}
 \epsfxsize=4.9cm
 \centerline{\epsfbox{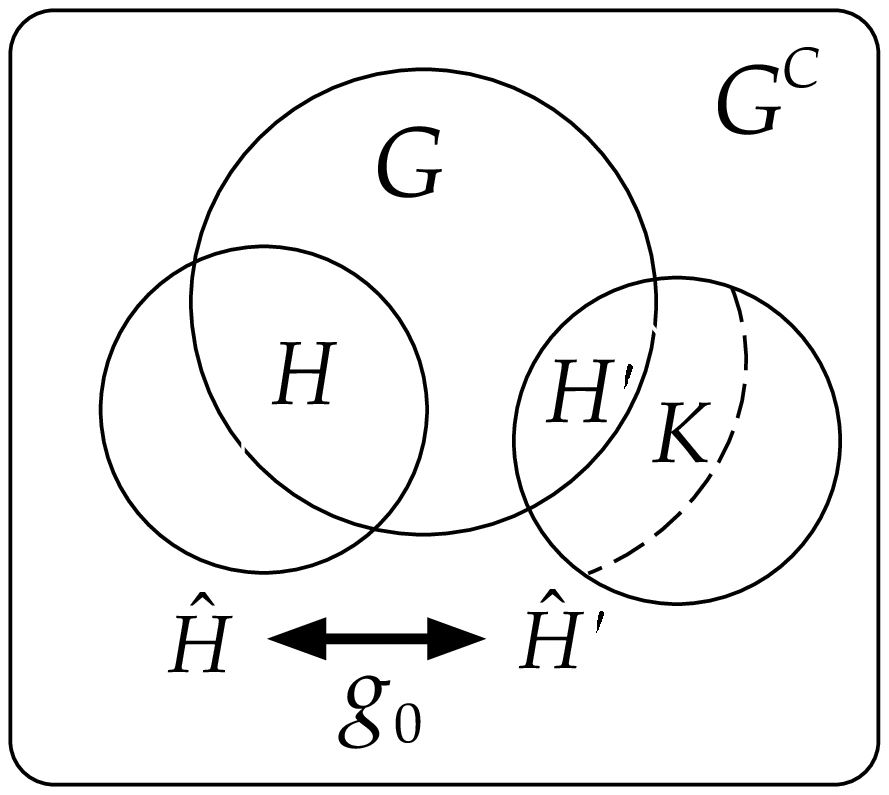}}
 \centerline{\mbox{Figure 2 : {\bf supersymmetric vacuum alignment}}}
  \begin{center}
  \begin{footnotesize}
The large circle indicates the group $G$. 
The small circles denote 
the complex subgroups ${\hat H}$ and ${\hat H}\pri$.
${\hat H}\pri$ is the transform of ${\hat H}$ by $g_0$.
The real subgroups $H$ or $H\pri$ are defined as 
intersections of $G$ and $\hat H$ or ${\hat H}\pri$.
$K$ is the image of $H$ by the $g_0$ transformation.
In general $H\pri$ is a subset of $K$.
  \end{footnotesize}
  \end{center}
\end{figure}
When the vacuum $\vec{v}$ is transformed 
by $g_0 \in G^{\bf C}$ to $\vecv = g_0 \vec{v}$,
the complex isotropy group $\hat H$ is transformed 
to $g_0 \hat H {g_0}^{-1}$.
They are isomorphic to each other.
On the other hand, the real isotropy group $H_v$ 
are not isomorphic to each other, 
since they are obtained by the equation 
\beq
 H_v = \hat H_v \cap G \;.
\eeq
Namely, the real isotropy $H$ at $v$ is transformed to 
$K = g_0 H {g_0}^{-1}$, 
but it is no longer included in $G$ (see Fig.~2).
Since the NG bosons parametrize the compact coset manifold $G/H_v$, 
their number changes at each point.
This means that $N_{\rm M}$ and $N_{\rm P}$ change. 
In fact, the Hermiticity of the broken generators changes 
at each point.
It is shown in Ref.~\cite{HN}  
how different compact cosets are embedded in the full manifold $M$. 
It is also shown that 
NG bosons are actually coupled to global $G$-currents there.
The number of the QNG bosons $N_{\rm Q} (=N_{\rm M})$ is
\beq
 N_{\rm Q} = \dim (G^{\bf C}/\hat H) - \dim (G/H_v) \;. \label{NQ}
\eeq
This number depends on the vacuum $v$ in the target space, 
since the dimension of $H_v$ also depends on.
Its range is shown in Sec.~3.4. 

It has been shown in Refs.~\cite{D-flat,GS} that 
there is a point such that 
$\hat H$ is reductive, 
namely $\hat H = H^{\bf C}$, $\dim {\cal B}=0$
and $N_{\rm P} = 0$, 
when the $G^{\bf C}$-orbit is a closed set.
We call such a point a symmetric point~\cite{HNOO,HN}.\footnote{
If the $G^{\bf C}$-orbit is not closed, 
we call the point where the unbroken symmetry is the largest 
as symmetric point.
In this sense, it is the point where 
Eq.~(\ref{Hhat-dec.}) is established.}
When all of the NG bosons belong to the mixed type, 
the realization is called the maximal realization.

\section{Moduli space of global symmetry}

\subsection{Algebraic geometry of invariants}
As the quotient space divided by $G^{\bf C}$ is parametrized by 
$G^{\bf C}$-invariant polynomials in $A^{G^{\bf C}}[V]$, 
the quotient space divided by $G$ is parametrized 
by $G$-invariant polynomials in $A^G[V]$.
Here, $A^G[V]$ is the ring of the $G$-invariant 
(but $not$ $G^{\bf C}$-invariant) polynomials.
We thus seek $G$-invariants composed by fundamental fields.
We denote the fundamental fields 
which belong to unitary representations $(\rho_i,V_i)$ 
as $\vec{\phi}_i$.
Their transformation laws under $G^{\bf C}$ are
\beq
 \vec{\phi}_i \gto \rho_i(g) \vec{\phi}_i \;,\;
 \vec{\phi}_i\dagg \gto \vec{\phi}_i\dagg \rho_i(g)\dagg 
 \;\;,\; g \in G^{\bf C}\;,
\eeq
where $\rho_i(g)$ are unitary matrices when $g \in G$, 
whereas they are $not$ unitary matrices 
when $g \in G^{\bf C}$.\footnote{
We will omit the symbol $\rho_i$ if not needed.}
Thus the moduli parameters can be chosen as
\beq
 \th_i = \vec{\phi}_i\dagg \vec{\phi}_i \in {\bf R}\;,\;
 \th_{i^*j} = \vec{\phi}_i\dagg \vec{\phi}_j \in {\bf C}\;. 
 \label{mod-para.}
\eeq
They are $G$-invariant, since we use the unitary representation.
Note that they are $not$ $G^{\bf C}$-invariant, 
since $\rho_i(g)\dagg\rho_i(g) \neq {\bf 1}$ for $g\in G^{\bf C}$.
The second invariants are possible 
when $\vec{\phi}_i$ and $\vec{\phi}_j$ 
are in the same representation of $G$.
Instead of complex numbers, we call real combinations 
$\th_{i^*j} + \th_{j^*i} \;,\; -i (\th_{i^*j} - \th_{j^*i}) $
moduli parameters in the rest of this paper.
The moduli parameters can be considered as 
being coordinates of the moduli space. 
We define the number of the $G$-invariant polynomials as $N(G)$, 
where
\beq
 N(G) \defeq \dim A^G [V] , 
\eeq
and we count them in the real dimension.\footnote{
$N(G)$ may change in the subspace of $M$ where 
some of the second type moduli parameters in Eq.~(\ref{mod-para.}) 
become real.}
Since the values of the moduli parameters are constant on each $G$-orbit, 
the moduli parameters can be considered to be a map from $G$-orbits to 
the moduli space ${\cal M}$, namely
\beq
 \pi : M \to {\cal M} = M/G \;.  \label{orbit-map}
\eeq
This kind of map is called an orbit map. 
(The orbit map in the type of $V/G$ 
has been discussed in Ref.~\cite{AS}.)
Conversely, each $G$-orbit is obtained by 
the inverse map $\pi^{-1}(p)$ from each point $p$ in the moduli space.  
A $G$-orbit is parametrized by NG bosons and is a coset space, $G/H_v$. 
In the generic region of the moduli space, 
a $G$-orbit has the maximal dimension, 
\beq
 \dim (G/H_{\rm g}) = \dim M - N_{\rm g}(G) \;,
 \label{dim.G/H}
\eeq
where the index g denotes the generic points in the moduli space.
At the singular points where the orbit shrinks, 
$\dim (G/H)$ takes smaller values than Eq.~(\ref{dim.G/H}). 

\bigskip
In general, 
the moduli parameters cannot take all values in ${\bf R}^{N(G)}$. 
The moduli space ${\cal M}$ is a subset of ${\bf R}^{N(G)}$, 
${\cal M} \subset {\bf R}^{N(G)}$, 
characterized by some inequalities between the moduli parameters such as
\beq
 r_C(\th_i,\th_{i^*j}) \geq 0 \;,\; C = 1,2, \cdots,  
\eeq
and the moduli space can be written as 
\beq
 {\cal M} = \{(\th_i,\th_{i^*j})\in {\bf R}^{N(G)}| 
                r_C(\th_i,\th_{i^*j}) \geq 0  \} \;.
\eeq
Although we do not give a general expression for these relations, 
we give some examples in the later section.

We decompose the moduli space ${\cal M}$ to some regions ${\cal M}_R$ as
\beq
 {\cal M} = \bigcup_R {\cal M}_R \;. \label{dec.mod.}
\eeq
Here, the label $R$ runs over ${\rm I},{\rm II},\cdots$. 
Each region is defined so that 
the unbroken symmetry at any point 
is isomorphic to each other by the $G^{\bf C}$ transformation, 
namely $H_p = g_0 H_q {g_0}^{-1}\,,\, g_0 \in G^{\bf C}$ 
for any two points $p,q \in {\cal M}_R$.
Note that they are not isomorphic to each other by $G$-action. 
We denote the conjugacy class of the real isotropy groups 
in region $R$ as 
\beq
 H_{(R)} = \{ H_p | p \in {\cal M}_R\} \;. \label{HR}
\eeq
In general, the dimensions of the various regions of the moduli space 
are different. 
(We describe a way to calculate them in Sec.~3.3. )
Thus the dimension of the moduli space is 
\beq
 \dim {\cal M} = {\rm sup}_R \dim {\cal M}_R \;. \label{sup-dim.}
\eeq
The dimension of the moduli space is given in Sec.~3.4.

\subsection{Effective Lagrangian}
The leading term of the effective Lagrangian is 
a nonlinear sigma model 
whose target manifold is a K\"{a}hler manifold, 
and can be written by a K\"{a}hler potential as~\cite{Zu,WB}
\beq
 {\cal L}_{\rm eff.} = \int d^4 \th K (\Ph,\Phd) \;.
\eeq
The most general $G$-invariant K\"{a}hler potential is 
written by an arbitrary function of the moduli parameters as
\beq 
 K = f(\th_i,\th_{i^*j}) \;. \label{Kahler}
\eeq
Since the function $f$ is a real function, 
the second type of variables appear in the form  
$\th_{i^*j} + \th_{j^*i}$ and $-i(\th_{i^*j} - \th_{j^*i})$.

Since the target space $M$ is a single $G^{\bf C}$-orbit, 
any point in $M$ is obtained by a $G^{\bf C}$-action 
from the vacuum $\vec{v}$. 
Hence, there is a remarkable relation 
between the fundamental fields $\vec{\phi}_i$ and the vacuum $\vec{v}_i$ as  
\beq
 \vec{\phi}_i|_{\rm F} = \xi \vec{v}_i \;. \label{rel.-phi-xi}
\eeq
Here, the subscript F means solutions of F-term constraints 
(corresponding to fixing the $G^{\bf C}$-invariants)
and $\xi$ is the representative of 
the complex coset $G^{\bf C}/\hat H$, written as
\beq
 \xi = \exp ( i \Ph^R Z_R ) \;,
\eeq
where $Z_R \in {\cal G}^{\bf C} - \hat{\cal H}$ 
are the complex broken generators and 
$\Ph^R$ are the NG chiral superfields, 
whose scalar component 
parametrize the K\"{a}hler coset manifold, $G^{\bf C}/\hat H$.
~From Eq.~(\ref{rel.-phi-xi}), 
the moduli parameters can be written as
\beq
 \th_i = \vec{\phi}_i\dagg \vec{\phi}_i|_{\rm F} 
     = \vec{v}_i\dagg \xi\dagg \xi \vec{v}_i \;\;,\;
 \th_{i^*j} = \vec{\phi}_i\dagg \vec{\phi}_j |_{\rm F} 
     = \vec{v}_i\dagg \xi\dagg \xi \vec{v}_j \;, \label{mod.para.}
\eeq 
where $\vec{v}_i$ satisfy the F-term constraints 
: $W\pri(\vec{v}_i)=0$. 
We thus find that this K\"{a}hler potential Eq.~(\ref{Kahler}) 
is just the BKMU's A-type Lagrangian~\cite{Le,BKMU}
(see also Ref~\cite{BL,Sh,KS,Ku}).\footnote{
BKMU found three types of K\"{a}hler potential~\cite{BKMU}.
In Eq.~(\ref{Kahler}), 
we have constructed the K\"{a}hler potential to be strictly invariant 
under $G$-transformations. 
If we require a K\"{a}hler potential 
that is not strictly $G$-invariant, 
but $G$-invariant up to a K\"{a}hler transformation, 
there may exist a K\"{a}hler potential 
corresponding to the BKMU's B-type K\"{a}hler potential. 
Since it has been known that 
the B-type appears in a pure realization~\cite{IKK} 
(the case when there exist only pure-type multiplets and 
there is no QNG boson), 
we think that the B-type can appear 
when there is any pure-type multiplet.
We do not know if 
it can appear when 
there exists the vacuum alignment and 
some pure-type multiplets turn to mixed-type multiplets. 
We do not obtain the C-type K\"{a}hler potential in our method.
}

If we choose the arbitrary function $f$ linear, 
$f(\th_i,\th_j,\cdots) = a \th_i + b \th_j + \cdots$, 
the space where fundamental fields $\vec{\phi}_i$ live 
is a flat linear space.
On the other hand, 
if we use variables of type $\th_{i^*j}$ or 
choose the arbitrary function $f$ nonlinear, 
the space where fundamental fields $\vec{\phi}_i$ live 
is no longer flat linear space. 

It was not known 
how many variables the arbitrary function of 
effective K\"{a}hler potential contains.
In our formalism, 
it is clear how many variables it contains; 
we show in a later section that 
the number coincides with the minimum number of QNG bosons 
and the dimension of moduli space.

It is known that the QNG bosons correspond to 
the non-compact directions 
of the target manifold~\cite{Le,BL,Sh,KS,HNOO,HN}. 
This is true even at points where 
the number of QNG bosons changes~\cite{HN}.  
Since the symmetry of the theory is compact group $G$, 
but not $G^{\bf C}$, 
it cannot control the non-compact directions.
This is why the QNG bosons bring arbitrariness to 
the K\"{a}hler potential 
of the effective Lagrangian~\cite{Le,BL,Sh,KS}.
In this paper it will become clear that 
the target space is the fibre bundle on the moduli space 
with the $G$-orbits as the fibre.
The arbitrary function can be interpreted as 
the freedom to change the size of 
the $G$-orbit at each point of the moduli space,  
since the derivatives of the arbitrary function 
is related to the decay constants of the NG bosons 
which parametrize the $G$-orbit~\cite{Sh,KS,HNOO,HN}.

\subsection{Geometry of K\"{a}hler coset manifolds}
In this subsection, 
we derive the moduli space by investigating  
the K\"{a}hler coset manifolds in detail. 
The target space of broken global symmetry, 
the K\"{a}hler coset manifold $M$, 
is a non-compact and non-homogeneous manifold.
The compact isometry group $G$ connects points in compact directions.
Two points apart to non-compact directions are connected 
by $G^{\bf C}$, but not $G$. 
Since the moduli space ${\cal M}$ is defined by $M/G$, 
only non-compact directions remain in ${\cal M}$. 
The target manifold $M$ is spanned by the broken generators, 
which can be decomposed as a direct sum of $H$-irreducible sectors.
In this subsection, we show that each irreducible sector 
comprising mixed-type generators corresponds to 
an independent non-compact direction.\footnote{
Kotcheff and Shore discussed the special case of 
the maximal realization at the symmetric point~\cite{KS} 
without any discussion about the moduli space.}

First of all, consider the vacuum $\vec{v}$ 
belonging to the $R$-th region of the moduli space ${\cal M}_R$, 
namely $\pi(\vec{v}) \in {\cal M}_R$, 
and transform it to another vacuum, $\vecv = g_0 \vec{v}$, 
where $g_0$ is the element of $G^{\bf C}$. 
To derive the moduli space, we need 
the $G^{\bf C}$-transformation modulo $G$-transformation. 
The element of $g_0 \in G^{\bf C}$ is divided into
\beq
 g_0 = \exp (i \th^R Z_R|_{\rm M}) \cdot
       \exp (i \th^R Z_R|_{\rm P}) \cdot \hat h \;, 
  \label{devided-g0}       
\eeq
where $Z_R|_{\rm M}$ are mixed-type generators and thus Hermitian; 
$Z_R|_{\rm P}$ are pure-type generators, and thus non-Hermitian; 
and $\hat h$ is an element of $\hat H_v$.
The last two components transform
$\vec{v}$ to points in the same $G$-orbit, 
since the last element does not move $\vec{v}$ and 
the second term can be absorbed by 
the some local $\hat H$ transformation as 
$\exp (i\th^R Z_R|_{\rm P}) \cdot \zeta  
= \exp (i a^i X_i)$, where $a^i$ are real and $X_i$ are 
some broken Hermitian generators. 
(Here, $a^i$ can be obtained 
by using the Baker-Campbell-Hausdorff formula. 
See Ref.~\cite{HN}.)
Therefore, we can omit them to obtain the moduli space.
Since mixed-type generators are Hermitian, 
they also belong to ${\cal G}-{\cal H}$ and 
are divided into $n_R$ $H$-irreducible sectors, 
since $h({\cal G}-{\cal H})h^{-1} \subset {\cal G}-{\cal H}$. 
Let the number of broken generators in the $i$-th sector be $m_i$ 
($i=1,\cdots,n_R$), 
then the complex broken generators are divided into 
\beq
 {\cal G}^{\bf C} - \hat{\cal H} 
 =\{ \{ Z_1^{(1)},\cdots, Z_{m_1}^{(1)} \}_{\rm M},\cdots, 
   \{ Z_1^{(n_R)},\cdots, Z_{m_n}^{(n_R)} \}_{\rm M} 
     || \mbox{P-type} \} ,
\eeq
and each sector is transformed by $h \in H_v$, as
\beq
 h Z_R^{(i)} h^{-1} = Z_S^{(i)} {\rho_i(h)^S}_R \;,
\eeq
where ${\rho_i(h)^S}_R$ are $m_i$-by-$m_i$ representation matrices. 
For later convenience, we write these sectors as
\beq
 {\bf m_1}_{\rm M} \oplus \cdots \oplus{\bf m_n}_{\rm M} 
 \oplus {\bf m_1}_{\rm P} \oplus \cdots \;,
\eeq
where the index M (P) denotes the mixed- (pure-) type sectors, 
and all components are $m_i$-dimensional 
irreducible representations of the unbroken symmetry, $H_v$.  

The independent transformations 
to non-compact directions, first element of 
Eq.~(\ref{devided-g0}), are at most  
\beq
 \vecv = \exp (i{\th^R}_{(i)} {Z^{(i)}}_R|_{\rm M}) \cdot \vec{v} \;,
  \label{transf.vac.}
\eeq
where all parameters ${\th^R}_{(i)}$ are pure imaginary.
This does not yet represent the independent non-compact directions. 
Since the new vacuum $\vecv$ do not preserve 
the unbroken symmetry $H_v$ at $\vec{v}$, 
it is transformed by $H_v$.
The $H_v$-orbit of the $\vec{v}\,\pri$ is
\beq
 h \cdot \vec{v}\,\pri 
&=& [ h \exp(i{\th^R}_{(i)} {Z^{(i)}}_R) h^{-1}] h \vec{v} \non
&=& \exp[i{\th^R}_{(i)} { (h {Z^{(i)}}_R h^{-1})}] \vec{v} \non
&=& \exp(i{\th^R}_{(i)} 
    \;{ {{\rho_i(h)}^S}_R {Z^{(i)}}_S} ) \vec{v} \non  
&=& \exp (i{{\th\pri}^R}_{(i)} {Z^{(i)}}_R) \cdot \vec{v}, 
\eeq
where we have used Eq.~(\ref{transf.vac.}) and defined 
\beq
 {\th\pri}^R}_{(i)} \defeq {{\rho_i(h)}^R}_S \,{\th^S}_{(i) \;.
\eeq
Since we have assumed that $Z^{(i)}_R$ belong to 
the same $H_v$-irreducible sector, 
$H_v$ acts transitively on 
the space of vacuum of the form of Eq.~(\ref{transf.vac.}). 

In the generic region, each $G$-orbit has the maximal dimension. 
Thus independent non-compact directions are 
parametrized by $G$-invariants.
Therefore, the number of independent non-compact directions
equals to the number of $G$-invariants, $N_{\rm g}(G)$. 
Since this is also calculated as 
the number of $H$-irreducible sectors of mixed-types, $n_{\rm g}$, 
we obtain a theorem concerning  
the number of independent non-compact directions 
in the $generic$ $region$, 
\beq
 \mbox{\bf theorem~1.}\hspace{0.5cm}
 N_{\rm g} (G) = n_{\rm g} \;. \label{theorem1}
\eeq
Note that the left-hand side is an algebraic geometrical quantity, 
whereas the right-hand side is a group-theoretical quantity. 

Although the theorem is valid in the generic region, 
it seems to also be valid in any region as  
\beq
 \mbox{\bf conjecture.}\hspace{0.5cm}
 N_R (G) = n_R \;. \label{conjecture}
\eeq
Here, the label $R$ denotes the regions of the moduli space 
defined in Eq.~(\ref{dec.mod.}).
Although we do not know any proof of this conjecture, 
we show that this is correct in many examples in the next section.
We have stated that $N_R(G)$ changes 
where some of the complex moduli parameters become real. 
In the last example in the next section, this indeed occurs, 
and $n_R$ also changes accordingly; 
we thus believe that the conjecture is true in any region.

\bigskip
In general, non-compact directions do not commute, 
namely $g_0 g_1 \neq g_1 g_0$, 
even if they are independent. 
There are commuting directions and non-commuting directions.

First of all, let us discuss the case 
when the generators in different sectors commute.
We transform $\vec{v}$ to $\vecv = g_0 \vec{v}$ 
by $g_0 = \exp (i\th^R Z_R^{(i)})$ in an $i$-th $H_v$-sector. 
We take $\th^R$ to be pure imaginary, 
$\th^R = i\tht^R$ ($\tht^R \in {\bf R}$).
In this case, 
only one linear combination, $\tht^R Z_R^{(i)}$, of 
the transformed $i$-th broken generators, 
$g_0 \{ Z_1^{(i)},\cdots, Z_{m_n}^{(i)} \} g_0^{-1}$ 
remains Hermitian, 
and the rest change to non-Hermitian, 
and therefore pure-type, generators. 
The other sectors remain unchanged 
(since we consider the case where 
the generator of $g_0$ commutes with them).  
Vacuum alignment can occur only in the $i$-th sector. 
There are $m_i$ NG and QNG bosons at the first vacuum $\vec{v}$, 
whereas there are the $2 m_i -1$ NG bosons and one QNG boson 
at the transformed vacuum $\vecv$, 
as far as the massless bosons in the $i$-th sector are concerned.
(Note that in a $H_v$-singlet sector with $m_i =1$, 
the number of NG and QNG bosons does not change, 
and vacuum alignment does not occur.)

In the case when some sectors do not commute, 
since the transformation by the $i$-th sector induces 
a transformation in the other sectors 
not commuting with the $i$-th sector,  
the generators of such sectors change to 
non-Hermitian, thus pure-type, generators.
Thus the vacuum alignments occur 
not only in the $i$-th sector, 
but also in the other sectors not commuting with the $i$-th sector.

When the vacuum $\vecv$ is rotated from the symmetric points 
by all non-commuting mixed-type sectors 
with pure imaginary parameters, 
it belongs to the generic region ${\cal M}_{\rm g}$ 
of the moduli space. 
The unbroken symmetry $H_{v\pri}$ becomes the minimum 
in the whole moduli space.
In such a region, 
all of the $H_{v\pri}$-sectors are singlets, 
since the vacuum alignment to smaller unbroken symmetry 
cannot occur, by definition.

\bigskip
We can calculate the dimension of 
each region of the moduli space.
Let us assume that the vacuum $\vec{v}$ belongs to 
the $R$-th region ${\cal M}_R$.
If we move the vacuum by the $H_v$-singlet sectors 
of the mixed-type broken generators, 
vacuum alignment does not occur.
By using the $i$-th sector of $m_i = 1$, 
we transform the vacuum $\vec{v}$ to $\vecv = g_0 \vec{v}$, 
where $g_0 = \exp (i\th Z^{(i)})$.
Since $Z^{(i)}$ is $H$-singlet, $H_{v\pri} = g_0 H_v g_0^{-1} = H_v$. 
Therefore, there is no vacuum alignment and 
the vacuum stays in the same region ${\cal M}_R$.

If there are several $H_{(R)}$-singlets at a point in the $R$-th region, 
there are the same number of directions 
which bring the vacuum to another vacuum 
in that region of the moduli space, ${\cal M}_R$. 
We thus obtain a theorem concerning 
the dimension of the $R$-th region of the moduli space,
\beq
 \mbox{\bf theorem~2.}\hspace{0.5cm}
 \dim {\cal M}_R = n_R({\bf 1}_{\rm M}) \;, \label{theorem2}
\eeq 
where we have defined the number of $H_{(R)}$-singlet sectors 
in the $R$-th region ${\cal M}_R$ as $n_R({\bf 1}_{\rm M})$, 
and ${\bf 1}_{\rm M}$ means the $H_{(R)}$-singlet of the mixed type.
The number of $H_{(R)}$-singlets is less than 
the number of all mixed-type sectors $n_R$, 
\beq
 n_R({\bf 1}_{\rm M}) \leq n_R \;. 
\eeq
The inequality is saturated in the generic region 
${\cal M}_{\rm g}$ of moduli space, 
since all of the $H_{\rm (g)}$-sectors of the mixed-type 
become singlets there, as discussed before. 
We thus obtain the following corollary: 
\beq
 \mbox{\bf corollary~1.}\hspace{0.5cm}
 \dim {\cal M}_{\rm g} = n_{\rm g}({\bf 1}_{\rm M}) = n_{\rm g} \;,
  \label{corollary1}
\eeq
where the index g means the generic region.
Since the dimension of the generic region is 
the largest in all moduli space, 
it is also the dimension of 
the moduli space itself from Eq.~(\ref{sup-dim.}).

\bigskip
We now comment on the geometry. 
The target space $M$ can be considered 
to be fibre bundle~\cite{HKLR}.\footnote{
Exactly speaking, not the full space $M$, 
but the space on each region of the moduli space, 
$\pi^{-1}({\cal M}_R)$, 
can be considered as fibre bundle, 
since the fibre at each region has different dimension.} 
The base space is the moduli space ${\cal M}$, 
the fibre is the $G$-orbit, $G/H_v$, and 
the structure group is $G$. 
The projection is the orbit map $\pi$ in Eq.~(\ref{orbit-map}).
The $G$-orbit shrinks somewhere on the moduli space. 
Consider some region ${\cal M}_R$ of moduli space and 
its boundary $\del {\cal M}_R$ (if it exists).
The boundary is another region ${\cal M}_{\del R}$. 
In general, the unbroken symmetry, $H_{(\del R)}$, in 
the boundary is larger than $H_{(R)}$ of the bulk,\footnote{
This phenomenon has been essentially discussed by Hull et al. 
in the third paper of Refs.~\cite{gauging}.} 
\beq
 H_{(\del R)} \supset H_{(R)}.
\eeq
$H_{(\del R)}$ is broken down to $H_{(R)}$ in the bulk 
and constitutes the $H_{(\del R)}$-orbit, $H_{(\del R)}/H_{(R)}$, 
in the $G$-orbit. 
Conversely, it shrinks in the boundary ${\cal M}_{\del R}$ and 
the unbroken symmetry, $H_{(R)}$, is enhanced to $H_{(\del R)}$.
This means that most of the NG bosons change to QNG bosons there. 
Namely, the NG-QNG change occurs.
(See for example, Fig.~5.)

\subsection{Dimension of moduli space}
In this subsection, 
we collect the obtained results 
concerning to the dimension of the moduli space. 
~From theorem~1 (Eq.~(\ref{theorem1})) 
and Eqs.~(\ref{corollary1}) and (\ref{dim.G/H}), 
we obtain the following corollary for the dimension of the moduli space: 
\beq
 \mbox{\bf corollary~2.}\hspace{0.5cm}
 \dim {\cal M} 
  &=& \dim {\cal M}_{\rm g}
  = n_{\rm g}({\bf 1}_{\rm M})
  = n_{\rm g} \non
 &=& N_{\rm g}(G)
  = \dim (G^{\bf C}/\hat H) - \dim(G/H_{\rm (g)}) \;.\label{corollary2}
\eeq
As shown in Sec.~3.2, 
this is also the number of variables in 
the arbitrary function in the effective K\"{a}hler potential.
It can also be stated as the number of decay constants 
of NG bosons.\footnote{
The authors of \cite{KS} have stated that this coincides with 
${\rm rank}\,G - {\rm rank}\,H$.
(See Appendix B in the second reference in Ref~\cite{KS}.)
But this is not correct in general. 
It is correct only in 
their example of the chiral symmetry breaking 
where ${\rm rank}\, H$ is unchanged.
(It is also true in the pure realization, 
which is the case when there is only the pure-type multiplets, 
and therefore $G^{\bf C}/\hat H \simeq G/H$~\cite{IKK}.
In such cases, 
although there is no vacuum vector and 
they have no linear model origin~\cite{Le,BL,KS}, 
there is no $G$-invariants namely $N(G)=0$, 
which coincides with ${\rm rank}\, G - {\rm rank}\, H = 0$.
It is also true when some Cartan generators are broken from 
the pure realization~\cite{BE}.)
To include other cases, 
even if we would modified it to 
${\rm rank}\,G - {\rm rank}\,H_{\rm (g)}$, 
it is correct only in the case 
where there is only one vacuum vector 
transforming in the irreducible representation 
(besides above cases).
See examples in the next section.}

\bigskip
We have shown that the number of QNG bosons changes 
even if we consider one theory.
It has been known 
that there must be at least one QNG boson in any low-energy theory 
with a fundamental theory origin~\cite{Le,BL,KS}. 
However the minimum number of QNG bosons was not known. 
In this paper, we have found it. 
The range of the number of QNG bosons, $N_{\rm Q}$, is calculated 
from Eqs.~(\ref{NQ}), (\ref{corollary2}) and (\ref{Hhat-dec.}) as 
\beq
 \dim (G^{\bf C}/\hat H) - \dim(G/H_{\rm (g)})  
  \leq & N_{\rm Q} & \leq 
 \dim (G^{\bf C}/\hat H) - \dim(G/H) \;,\non
 \dim {\cal M} \leq & N_{\rm Q} & 
  \leq \dim (G/H) - \dim B  \;.
\eeq
The left-hand inequality is saturated at the generic points,
where the QNG bosons are tangent vectors of the moduli space as
\beq
 \pi_* (\mbox{QNG}) = T_p \,{\cal M} \;,\; p \in {\cal M}_{\rm g} \;,
\eeq
where $\pi_*$ is a differential map of orbit map (\ref{orbit-map}); 
the right inequality, however, is saturated 
at the most singular points in the moduli space 
corresponding to the symmetric points in the target space.
(For closed $G^{\bf C}$-orbits, 
there is no Borel algebra (${\cal B}=0$) and 
the complex isotropy $\hat H$ is reductive at symmetric points. 
The maximal realization occurs and 
there are equal numbers of QNG and NG bosons.)

\section{Examples}
In this section, 
we give some examples, 
demonstrate the theorems and corollaries obtained in the last section 
and confirm that the conjecture is correct. 
Although we treat only the group $O(N)$ and $SU(N)$, 
the extension to another group is straightforward.
The first several subsections are 
devoted to fundamental representations, 
and the last two subsections to adjoint representations.
Rapid readers should read only Sections~4.1,~4.3 and 4.4.

\subsection{Example of $\dim {\cal M} = 1$ (closed set)}
Example~1) $O(N)$ with $\vec{\phi} \in {\bf N}$

Since this is the simplest example, 
we investigate this in detail.
The physical result, such as low-energy theorems, 
are considered in Ref.~\cite{HN}. 

First of all, we define the generators of the group $O(N)$ as
\beq
 {(T_{ij})^k}_l 
   = \1{i}({\delta_i}^k \delta_{jl} - {\delta_j}^k \delta_{il}) . 
\eeq 
If we consider only the real group $G=O(N)$, 
the fundamental fields $\vec{\phi}$ live in 
the real space ${\bf R}^N$, 
since it is a real representation.
However, we need the complex extension of $G$, $G^{\bf C}$, 
and they live in the complexified space ${\bf C}^N$.
There is only one $G^{\bf C}$-singlet, $\vec{\phi}\,^2$, 
namely $N(G^{\bf C}) = 1$.
We consider a $G$-invariant superpotential,
\beq
 W = g \phi_0 (\vec{\phi}\,^2 - f^2)  \;,
\eeq
where $\phi_0$ is a singlet of $G$ and 
a non-dynamical auxiliary field. 
Here, $f$ is a real nonzero constant.\footnote{
There are four kinds of $G^{\bf C}$-orbits.
They are classified by 
the value of the $G^{\bf C}$-invariant polynomial $\vec{\phi}^2$,
namely $f$.
We now consider the case when $f$ is real and nonzero 
as shown in Fig.~3.
There is another orbit that $f$ is pure imaginary and nonzero.
Other two types are cases that $f$ is zero.
One is the case when $\vec{v}$ itself is zero.
Another is the case when $\vec{v} \neq 0$.
Only the last one is an open set.
We do not consider such orbits, 
since it has no real submanifold corresponding to 
the NG manifold without QNG bosons.}
Actually, the superpotential $W$ is $G^{\bf C}$-invariant, 
since it includes only chiral multiplets.
If we take the limit $g \to {\rm infinity}$ and decouple $\phi_0$, 
we obtain the F-term constraint, 
\beq
 \vec{\phi}\,^2 - f^2 =0  \;.\label{F-O(N)} 
\eeq
(We have obtained it from the equation of motion of $\phi_0$.)
The obtained space is the $G^{\bf C}$-orbit with  
the complex dimension 
$N_{\Ph} = \dim_{\bf C}V - N(G^{\bf C})=2N-1$.
It is the target space of the effective 
nonlinear sigma model.

The vacuum $\vec{v} = <\vec{\phi}>$ at the target space 
can be transformed to
\beq
 \vec{v} = \pmatrix{0 \cr \vdots \cr 0 \cr { v}\cr}  
\eeq
by the $G^{\bf C}$-action, 
where $v$ is a real positive constant equal to $f$: $v=f$. 
We call this point a symmetric point and 
this region is called region~I.
The unbroken symmetry is $H=O(N-1)$ and 
the real broken generators are
\beq
 X_i = T_{Ni} 
 = \left(
   \begin{array}{ccc|c}
              &          &          & \LARGE{0}  \\
              &\LARGE{0} &          & i         \\ 
              &          &          & \LARGE{0}  \\ \hline
   \LARGE{0}  &  -i      &\LARGE{0} & 0  
   \end{array}
   \right) \in {\cal G - H} \;\;\;(i = 1,\cdots ,N-1) \;. 
 \label{br.O(N)}
\eeq
Since the NG bosons are generated by these generators, 
the real target space is $G/H = O(N)/O(N-1)$ 
embedded in the full target space $M$.
The complexification does not change the situation and 
the complex broken and unbroken generators coincide with 
those of real generators: 
\beq
 Z_R = X_i \in {\cal G}^{\bf C} - \hat {\cal H} \;,\;
 K_M = H_a \in \hat {\cal H} \;.
\eeq
Since they are Hermitian, 
the maximal realization occurs, namely 
$M \simeq G^{\bf C}/\hat H = O(N)^{\bf C}/O(N-1)^{\bf C}$.
Thus the numbers of NG and QNG bosons are both $N-1$.
Note that the broken generators 
belong to a single representation ${\bf N-1}$ of $H = O(N-1)$.
We denote this as $({\bf N-1})_{\rm M}$, 
where subscript M denotes a mixed type
(see the first line of Table~1).
Thus the number of $H_{\rm (I)}$-irreducible sectors of 
the mixed-type generators in the region~I is $n_{\rm I}=1$.
Since there is no $H_{\rm (I)}$-singlet sector, 
the dimension of region~I is 
$\dim {\cal M}_{\rm I}=0$ from theorem~2. 
Since there is only one $G$-invariant $|\vec{\phi}|^2$, 
we have also verified the conjecture in region~I as 
$N_{\rm I}(G) = n_{\rm I}=1$.

We transform the vacuum to another one, 
which belong to region~II, 
by $G^{\bf C}$-action, as 
\beq
 { \vec{v}} = \pmatrix{0 \cr \vdots \cr 0 \cr { v}\cr}  
 \stackrel{ g_0\in G^{\bf C}}{\to} 
 { \vec{v}\,\pri  = g_0 \vec{v}} 
 = \pmatrix{0 \cr \vdots \cr 0 \cr -i v \sinh \tht \cr  v \cosh \tht \cr} 
 \defeq \pmatrix{0 \cr \vdots \cr 0 \cr{ \alpha} \cr { \beta}\cr} \;,
\eeq
where we have put $g_0$ as
\beq
 g_0 
 &=& \exp(i \th X_{N-1}) \non 
 &=& \left(
   \begin{array}{c|cc}               
               \LARGE{1} &\LARGE{0} &          \\ \hline
                         & \cos \th & -\sin\th \\ 
               \LARGE{0} & \sin \th &  \cos\th  
   \end{array}
   \right)
 = \left(
   \begin{array}{c|cc}
               \LARGE{1} & \LARGE{0}   &             \\ \hline
                         & \cosh \tht  & -i\sinh\tht \\ 
               \LARGE{0} & i\sinh \tht &  \cosh\tht  
   \end{array}
   \right) \;.
\eeq
Since the other broken generators belong to 
the same $H_{\rm (I)}$ representation, 
they do not induce the independent non-compact directions, 
as shown in Sec.~3.3.
The real unbroken symmetry is $H\pri = O(N-2)$ and 
the real broken generators are $X_i$ in Eq.~(\ref{br.O(N)}) and  
\beq
 {X_i}\pri 
 = T_{N-1,i\pri} 
 = \left(
   \begin{array}{ccc|cc}
               &          &          &\LARGE{0} &   \\
               &\LARGE{0} &          & i        & 0 \\ 
               &          &          &\LARGE{0} &   \\ \hline
               &  -i      &          & 0        & 0 \\ 
     \LARGE{0} &  0       &\LARGE{0} & 0        & 0  
   \end{array}
   \right)  \;\;\;(i\pri = 1,\cdots ,N-2) \;.
\eeq
The real target manifold is $G/H\pri = O(N)/O(N-2)$, 
generated by $X_i$ and ${X_i}\pri$,
which has more dimensions than $G/H$ (see Fig.~4).
\begin{figure}
  \epsfxsize=8cm
  \centerline{\epsfbox{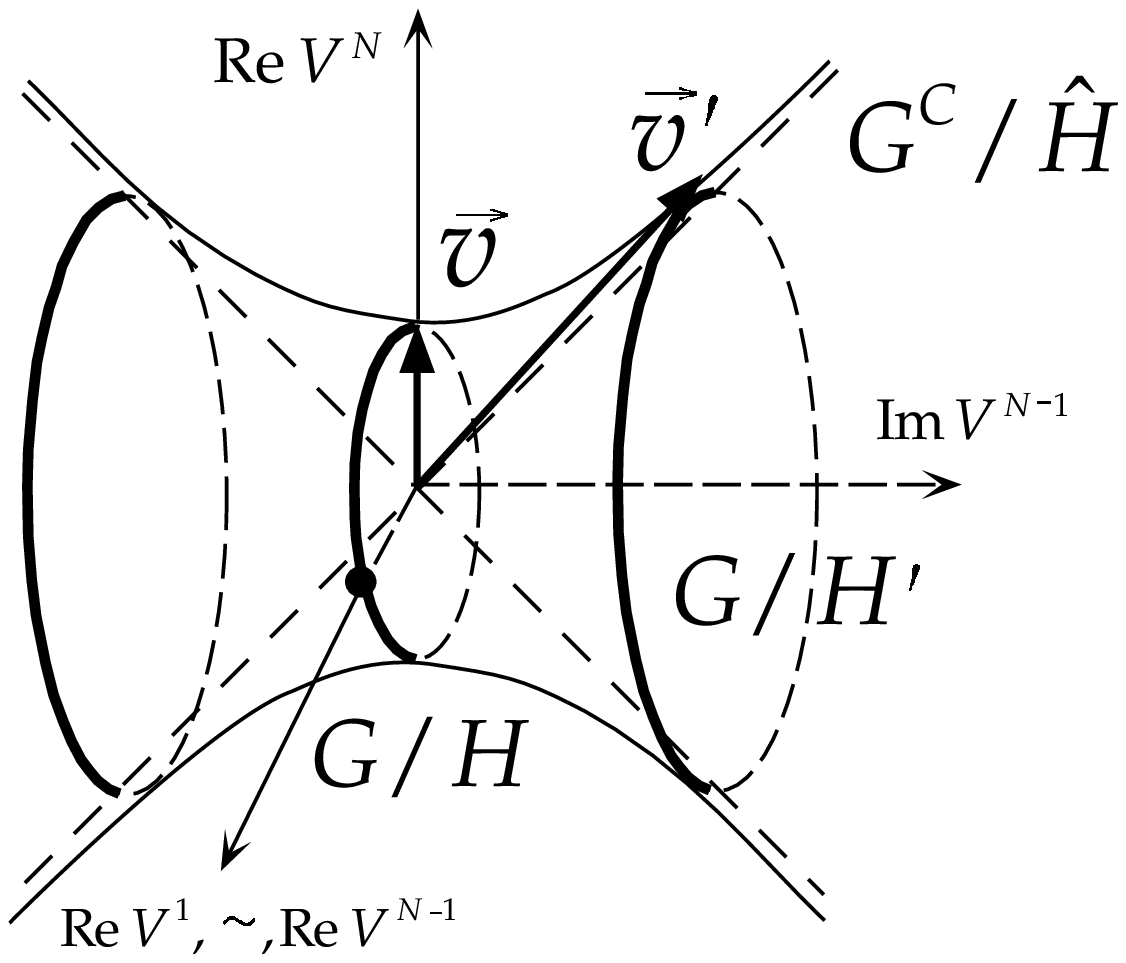}}
  \centerline{\mbox{Figure 3 : {\bf target space of the $O(N)$ model 1}}}

  \epsfxsize=8cm
  \centerline{\epsfbox{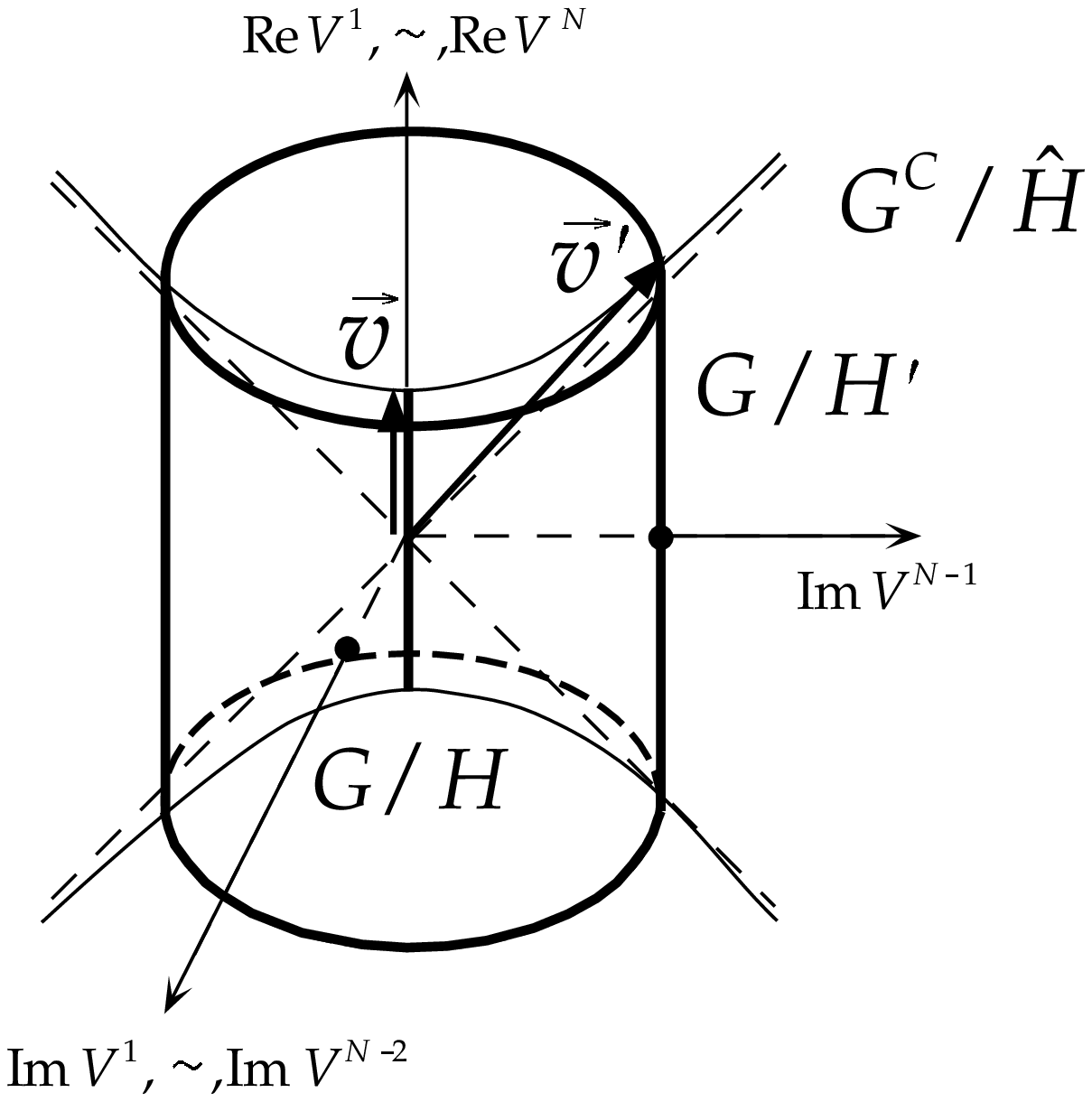}}
  \centerline{\mbox{Figure 4 : {\bf target space of the $O(N)$ model 2}}}
\end{figure}

The complex broken and unbroken generators at 
the non-symmetric points are obtained as
\beq
 {Z_R}\pri 
 &=& g_0 Z_R g_0^{-1} \non
 &=& \cases {g_0 X_i g_0^{-1}
             = { {\alpha \over v} {X_i}\pri 
             + {\beta \over v} X_i}  \;:\, \mbox{Pure-type} \cr
             g_0 X_{N-1} g_0^{-1} = { X_{N-1}}\;:\,\mbox{Mixed-type}
             }  
  \;\;\in {\cal G}^{\bf C}- \hat{\cal H}\pri \;, \label{ge.mixZ}
 \\
 {K_M}\pri 
 &=& g_0 K_M g_0^{-1} \non
 &=& \cases {g_0 X_i\pri g_0^{-1}
             = { {\beta \over v} {X_i}\pri 
             - {\alpha \over v} X_i}  \in \tilde{\cal B}\pri\cr
             g_0 {H_a}\pri g_0^{-1} = {H_a}\pri \in {\cal H}\pri
             } 
 \;\; \in \hat{\cal H}\pri  \;.\label{ge.mix}
\eeq
We write these as
\beq
 {\cal G}^{\bf C} - \hat {\cal H} 
 = \left(
   \begin{array}{ccc|cc}
        &          &     &             & \\
        &\LARGE{0} &     &\; \LARGE{P} & \\ 
        &          &     &             & \\ \hline
        &          &     & 0           & M  \\
    \;\;&\LARGE{P} &\;\; & M           & 0 
   \end{array}
   \right)  \;,\;
 \hat {\cal H} 
 = \left(
   \begin{array}{ccc|c}
              &                        & &  \\
 &\LARGE{\hat{\cal H}^{\prime {\bf C}}}& &\;\LARGE{\tilde{\cal B}\pri}\;\\ 
              &                        & &  \\ \hline
              &                        & &  \\
            &\LARGE{\tilde{\cal B}\pri}& & \;\LARGE{0} \;
   \end{array}
   \right)  \;,
\eeq
where P and M denote the pure- and mixed-type generators and 
$\tilde{\cal B}\pri$ represents the non-Hermitian, 
but not Borel, generators. 
The unbroken generators also include Hermitian and 
non-Hermitian generators.
The Hermitian generators coincide 
with real symmetry ${\cal H}\pri$,
whereas non-Hermitian one concernes  
the newly emerged NG bosons.
(It is shown in Ref.~\cite{HN}  
how different compact manifolds are embedded 
in the full manifold.)
At the non-symmetric point,  
only one of complex broken generators, $X_{N-1}$, 
is Hermitian, and thus a mixed-type generator, 
on the other hand, 
since the others are non-Hermitian, 
and they are the pure-type generators.
Thus, the numbers of the NG chiral multiplets are 
$N_{\rm M} = 1$ and $N_{\rm P} = N-2$.
The emergence of pure-type generators is 
the result of QNG-NG change (see Fig.~5).
There are $2N-3$ NG bosons and only one QNG boson.
The mixed one belongs to a single representation 
${\bf 1}_{\rm M}$ of $H\pri = O(N-2)$, and, 
the others belong to $({\bf N-2})_{\rm P}$ (see Table~1).
Since the number of the singlet is $n_{\rm II}({\bf 1}_{\rm M})=1$, 
the dimension of region~II is 
$\dim {\cal M}_{\rm II}=1$ from theorem~2. 
Since region II is the generic region, of course, 
theorem~1 is ture: $N_{\rm II}(G) = n_{\rm II}=1$.
\begin{figure}
  \epsfxsize=6cm
  \centerline{\epsfbox{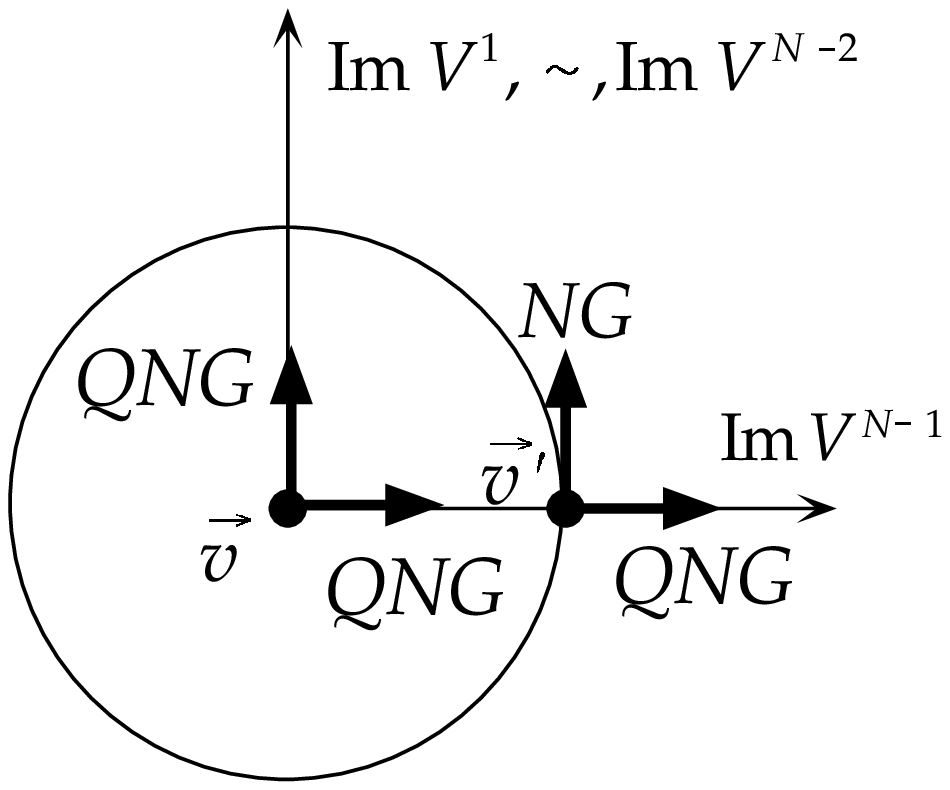}}
  \centerline{\mbox{Figure 5 : {\bf QNG-NG change}}}
\end{figure}

We have shown that there are two kinds of vacua in this model.
One is a symmetric point in region~I; 
the other is a non-symmetric point in region~II.
In the target manifold $M$, 
there are the same vacua generated by $G$, 
corresponding to NG directions.
Therefore, it is useful to see distinct vacua 
that we define moduli space as 
a quotient space divided by the symmetry $G$ : ${\cal M} = M/G$.
We define the moduli parameter as~\footnote{
In examples we define the square root of first types of 
Eq.~(\ref{mod-para.}) as the moduli parameters.}
\beq
 \th_1 \defeq |\vec{\phi}|_{\rm F} = |\xi\vec{v}| \;,
\eeq
where F denotes the F-term constraint Eq.~(\ref{F-O(N)}) 
and $\xi$ is the representative of 
the complex coset, $G^{\bf C}/\hat H$. 
This parameter has a minimum as shown in Figs.~3 and~4.
The moduli space of this model can be written as 
\beq
 {\cal M} = \{ \th_1 \in {\bf R} | \th_1 \geq f \} \;.
\eeq 
This is a closed set and has two phases corresponding to 
the symmetric point (region I) and 
the non-symmetric point (region II) as, can be  seen in Fig.~6.
\begin{figure}
  \epsfxsize=4.9cm
  \centerline{\epsfbox{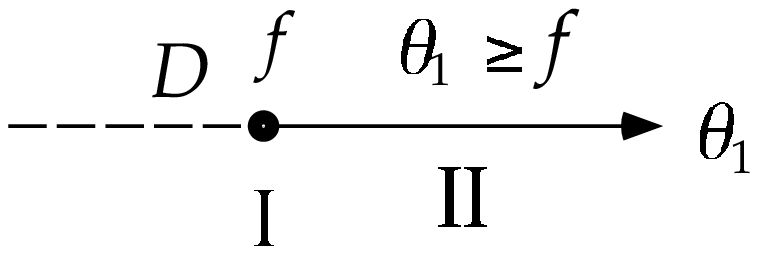}}
  \centerline{\mbox{Figure 6 : {\bf moduli space of $O(N)$ with ${\bf N}$}}}
\end{figure}
\begin{table}
\caption{\bf phases of $O(N)$ with ${\bf N}$}
\begin{center}
\begin{tabular}{|c|c|c|c|c|c|c|c|c|c|}
 \noalign{\hrule height0.8pt}
 $R$ & $H_{(R)}$ & $ N_{\rm M}$ & $N_{\rm P}$ & NG & QNG &
 $H_{(R)}$-sector & {\footnotesize $\dim {\cal M}_R$} & $n_R$ & $N_R(G)$ \\
 \hline
 \noalign{\hrule height0.2pt}
  I & $O(N-1)$ & $N-1$ & $0$   & $N-1$  & $N-1$ &
      $({\bf N-1})_{\rm M}$ & $0$ & $1$ & $1$   \\  
 II & $O(N-2)$ & $1$   & $N-2$ & $2N-3$ & $1$  & 
      $({\bf N-2})_{\rm P} \oplus {\bf 1}_{\rm M}$ & $1$ & $1$ & $1$  \\
 \noalign{\hrule height0.8pt}
 \end{tabular}
 \end{center}
\end{table}

The K\"{a}hler potential of 
the low-energy effective Lagrangian describing  
the behavior of the NG and QNG bosons 
is written by using moduli parameter as
\beq
 K = f({\th_1}^2) 
   = f(\vec{\phi}\,\dagg \vec{\phi})|_{\vec\phi^2 = f^2}
   = f(\vec{v}\,\dagg \xi\dagg \xi \vec{v}) \;,
\eeq
with a constraint $\vec{v}^{\,2} = f^2$.
The physical consequences, such as 
the low-energy theorems of the scattering amplitudes of 
the NG and QNG bosons, are discussed in Ref.~\cite{HNOO,HN}.

\subsection{Example of $\dim {\cal M} = 1$ (open set)}
Example~2) $SU(N)$ with $\vec{\phi} \in {\bf N}$

In the last example, 
the maximal realization occurs at the symmetric point, 
and $\hat H$ does not include the Borel-type algebra.
This example contains it. (See Ref.~\cite{LRM}.)

Since there is only the field in the complex representation
$\vec{\phi} \in {\bf N}$ (no conjugate representation), 
we cannot compose the $G^{\bf C}$ singlet 
in the superpotential.\footnote{
Mathematically, we can say that 
the coset $SU(N)/SU(N-1)$ cannot be embedded to 
the complex plane $(\vec{\phi} \in) {\bf C}^N$,  
since $\dim(SU(N)/SU(N-1)) = 2N-1$ and 
the constraints have at least two dimensions.}
Since $N(G^{\bf C}) = 0$,
the complex dimension of the target space is 
$N_{\Ph} = \dim_{\bf C}V - N(G^{\bf C})=N$.

As in Example~1, 
the vacuum $\vec{v} = <\vec{\phi}>$ can be transformed to
\beq
 \vec{v} = \pmatrix{0 \cr \vdots \cr 0 \cr { v}\cr}  
\eeq
by the $G^{\bf C}$-action, 
where $v$ is a real positive $arbitrary$ constant.
In this case, 
there is no other $G^{\bf C}$-orbit, 
except for the orbit $\vec{\phi} =0$.

The real unbroken symmetry is $H=SU(N-1)$ and 
the number of the NG bosons is $2N-1$. 
(The number of the QNG bosons should be one.)
We define $2(N-1)$ complex (non-Hermitian) generators, 
\beq
 X_i^-
 = \left(
   \begin{array}{ccc|c}
              &          &          & \LARGE{0} \\
              &\LARGE{0} &          &   0      \\ 
              &          &          & \LARGE{0} \\ \hline
    \LARGE{0} &   1      &\LARGE{0} & 0  
   \end{array}
   \right) \;,\;
 X_i^+
 = \left(
   \begin{array}{ccc|c}
              &          &          & \LARGE{0}  \\
              &\LARGE{0} &          & 1         \\ 
              &          &          & \LARGE{0}  \\ \hline
    \LARGE{0} &    0     &\LARGE{0} & 0  
   \end{array}
   \right) \;, \label{X+-}
\eeq
where $i=1,\cdots,N-1$.
The complex unbroken symmetry is 
\beq
 \hat{\cal H} 
 = \left(
   \begin{array}{ccc|c}
      &                         &  &      \\
      &\LARGE{{\cal H}^{\bf C}} &  &\LARGE{0}  \\ 
      &                         &  &      \\ \hline
      &\LARGE{\cal B}           &  & 0  
   \end{array}
   \right)   ,
\eeq
where $N-1$ generators ${\cal B}=\{X_i^-\}$ is the Borel generators.
Therefore, the complex unbroken generator $\hat H$ is not reductive, 
even at the symmetric point in this case.
The complex broken generators are
\beq
 {\cal G}^{\bf C} -\hat{\cal H} 
 = \left(
   \begin{array}{ccc|c}
          M &           & &      \\
            &\ddots     & &\LARGE{P}  \\ 
            &           &M&      \\ \hline
            &\LARGE{0}  & & M  
   \end{array}
   \right)   . \label{ex2}
\eeq
They comprise $N-1$ non-Hermitian pure-type generators 
$\{X_i^+\}$ (denoted P) 
and one Hermitian mixed-type generator (denoted M), 
which is a diagonal one: $N_{\rm P} = N-1$ and $N_{\rm M}=1$.
Thus, as noted above, there are $2N-1$ NG bosons and 
one QNG boson (see Table~2).

In this example, the $G^{\bf C}$-orbit is an open set.
We transform the vacuum as
\beq
 \vec{v} \to \vecv = g_0 \vec{v} \;,
\eeq 
where $g_0 = \exp(i\th X^{\rm diag.})$. 
Here, $X^{\rm diag.} \sim {\rm diag.}(1,\cdots,1,-(N-1))$, 
denoted M in (\ref{ex2}).
The angle is taken, to be pure imaginary, $\th = i \tht$. 
If we take the limit $\tht \to {\rm infinity}$, 
it reaches $\vec{\phi}= 0$.
Since the origin is omitted (it is an another orbit), 
the orbit is an open set.

If we define the moduli parameter as
\beq
 \th_1 \defeq |\vec{\phi}| = |\xi \vec{v}| \;,
\eeq
the moduli space is (see Fig.~7)
\beq
 {\cal M} = \{ \th_1 \in {\bf R} | \th_1 > 0 \} \;.
\eeq
The moduli space is also an open set.
\begin{figure}
  \epsfxsize=3.2cm
  \centerline{\epsfbox{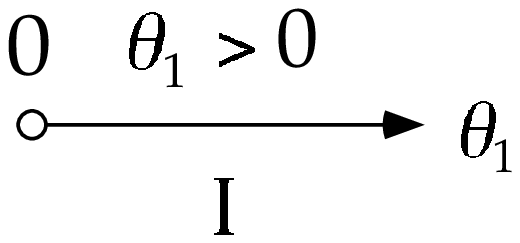}}
  \centerline{\mbox{Figure 7 : {\bf moduli space of $SU(N)$ with ${\bf N}$}}}
\end{figure}
\begin{table}
\caption{\bf phase of $SU(N)$ with ${\bf N}$}
\begin{center}
\begin{tabular}{|c|c|c|c|c|c|c|c|c|c|}
 \noalign{\hrule height0.8pt}
 $R$ & $H_{(R)}$ & $ N_{\rm M}$ & $N_{\rm P}$ & NG & QNG &
 $H_{(R)}$-sector & {\footnotesize $\dim {\cal M}_R$} & $n_R$ & $N_R(G)$ \\
 \hline
 \noalign{\hrule height0.2pt}
 I & $SU(N-1)$ & $1$ & $N-1$ & $2N-1$ & $1$ &
     $({\bf N-1})_{\rm P} \oplus {\bf 1}_{\rm M} $ & $1$ & $1$ & $1$ \\ 
 \noalign{\hrule height0.8pt}
 \end{tabular}
 \end{center}
\end{table}

There is only one region in this model.
We can find that fact soon. 
The dimension of the $G^{\bf C}$-orbit 
coincides with the dimension $V$, 
since there is no $G^{\bf C}$-invariant.
Since the origin is always omitted, 
it should be an open set, and there is only one phase.

\subsection{Example of $\dim {\cal M} = 2$}
Example~3) $SU(N)$ with 
$\vec{\phi} \in {\bf N}\;,\; \vec{\tilde{\phi}} \in \bar{\bf N}$

This example is seen in Ref.~\cite{LRM}.
The last two examples contain only 
one representation vector.
In such cases, the moduli space has one dimension, 
since the vector belongs to an irreducible representation.
In this example, we introduce two representation fields.
One belongs to a fundamental representation and 
the other to an anti-fundamental representation, 
transforming as~\footnote{
Actually there exist two kind of the anti-fundamental representation.
One is given here, whereas the other is transformed as 
$\vec{\tilde{\phi}}^T \to \vec{\tilde{\phi}}^T \cdot g\dagg$.
Although both coincide in the transformation of $G$, 
they are distinct in the transformation of $G^{\bf C}$.
We do not use this one, because we cannot construct 
$G^{\bf C}$-invariants.}
\beq
 \vec{\phi} \to g \cdot \vec{\phi} \;,\;
 \vec{\tilde{\phi}}^T \to \vec{\tilde{\phi}}^T \cdot g^{-1} \;.
\eeq
Since there is one $G^{\bf C}$-invariant,
$\vec{\tilde{\phi}} \cdot \vec{\phi}$, $N(G^{\bf C}) = 1$.
We construct the $G$-invariant superpotential 
with a non-dynamical singlet $\phi_0$, 
\beq
 W = g \phi_0 (\vec{\tilde{\phi}} \cdot \vec{\phi} - f^2) \;,
\eeq
which is also $G^{\bf C}$-invariant.
As stated in Example~1,
we obtain the F-term constraint, 
$\vec{\tilde{\phi}} \cdot \vec{\phi} - f^2 =0$.
The complex dimension of the target space is 
$N_{\Ph} = \dim_{\bf C}V - N(G^{\bf C}) = 2N-1$.

There exist two $G$-singlets: $N(G) = 2$, 
\beq
 \vec{\phi}\dagg \vec{\phi}\;,\; 
 \vecpht\dagg \vecpht \in {\bf R} \;.
\eeq
Thus as seen in later, 
the minimum number of the QNG and the dimension of 
the moduli space should be two from corollary~2 
(Eq.~(\ref{corollary2})).

\bigskip
The vacuum 
$\vec{v} = <\vec{\phi}> \;,\; \vec{\tilde{v}} = <\vec{\tilde{\phi}}>$
can be transformed by $G^{\bf C}$ to 
\beq
 \vec{v} = \pmatrix{0 \cr \vdots \cr 0 \cr { v}\cr}\;,\;
 \vec{\tilde{v}}^T = (0,\cdots,0,\tilde{v}) \;,\;  
\eeq
where
\beq
 v \vt = f^2 \;,\; v,\vt \in {\bf R} > 0 \mbox{ ( or $<0$)}.
\eeq
Since the case when the product $v\vt$ is negative 
corresponds to another $G^{\bf C}$-orbit, 
we omit such a case, as denoted in Example~1.
We call this point a symmetric point, 
where $\vec{v} \propto \vec{\tilde{v}}$.
It belongs to region~I in the moduli space.
The breaking pattern is $G=SU(N) \to H_{\rm (I)}=SU(N-1)$.
Since complexification does not change the situation,
it is a maximal realization. 
Thus the target space is
$G^{\bf C}/\hat H = SL(N,{\bf C})/SL(N-1,{\bf C})$, 
and there are $N-1$ NG bosons and $N-1$ QNG bosons.

We can transform the vacuum to 
the non-symmetric point belonging to region II, 
where $\vec{v} \propto\hspace{-0.35cm}/ \;\vec{\tilde{v}}$.
We choose $g_0 \in G^{\bf C}$ as
\beq
 g_0 = \exp(i \th X_{N-1}^-) 
 = \left(
   \begin{array}{c|cc}
               \LARGE{1} & \LARGE{0}  &   \\ \hline
                         & 1          & 0 \\ 
               \LARGE{0} &-\alpha/\vt & 1  
   \end{array}
   \right) \;,\;
g_0 ^{-1} 
 = \left(
   \begin{array}{c|cc}
               \LARGE{1} & \LARGE{0}  &   \\ \hline
                         & 1          & 0 \\ 
               \LARGE{0} & \alpha/\vt & 1  
   \end{array}
   \right)  \;, 
\eeq
where $X_{N-1}^-$ is defined in Eq.~(\ref{X+-}) and 
we have put $e^{\th} = -\alpha/\vt > 0\;(\alpha \in {\bf R} < 0)$.
The transformation by $g_0$ is 
\beq
 \vec{v} \to \vecv = g_0 \vec{v} =  \vec{v} \;,\;
 \vec{\vt}^T \to \vec{\vt}^{\,\prime T} = 
 \vec{\vt}^T g_0^{-1} = (0,\cdots,\alpha, \vt)  \;.
\eeq
Note that this transformation does not change $\vec{v}$.
The breaking pattern is 
$G = SU(N) \to H_{\rm (II)} = SU(N-2)$.
The number of NG bosons is $4N-4$ 
and the number of QNG bosons is $2$.
Although we can show these counting schemes in a generator level 
as in Eqs.~(\ref{ge.mixZ}) and (\ref{ge.mix}) in Example~1, 
we do not repeat it, since it is straightforward.  
The results are given in Table~3.

Since the length of the vectors are
\beq
 |\vecv|^2 = |\vec{v}|^2 = v^2 \;,\; 
 |\vecvt|^2 = |\vecvt\,\pri|^2 = \vt^2 + \alpha^2 \;,
\eeq 
the $\vec{v}$-fixing action induced by the $X_{N-1}^-$ 
decreases the length of $\vt$ as 
$|\vecvt\,\pri|^2 \geq \vt^2$.
There is an another independent transformation 
induced by $X_{N-1}^+$, defined in Eq.~(\ref{X+-}).
This time it fixes $\vec{\vt}$ and decreases
the length of $\vec{v}$ : $|\vecv|^2 \geq v^2$ 

If we define moduli parameters as
\beq
 \th_1 \defeq |\vec{\phi}|_{\rm F} = |\xi\vec{v}| \;,\; 
 \th_2 \defeq |\vec{\pht}|_{\rm F} = |(\xi^{-1})^T\vecvt| \;,
\eeq
the moduli space is (see Fig.~8)
\beq
 {\cal M} = \{ (\th_1 , \th_2 )\in {\bf R}^2 | \th_1 \th_2 \geq f^2 \}\;,
\eeq
~from the argument above. This is a closed set.
\begin{figure}
  \epsfxsize=4.9cm
  \centerline{\epsfbox{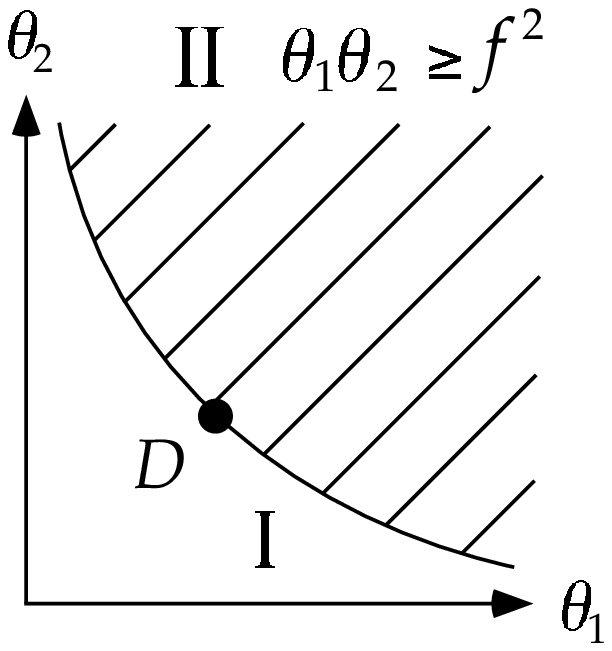}}
  \centerline{
  \mbox{Figure 8 : {\bf moduli space of $SU(N)$ with ${\bf N,\bar{N}}$}}}
\end{figure}
\begin{footnotesize}
\begin{table}
\caption{\bf phases of $SU(N)$ with ${\bf N,\bar{N}}$}
\begin{center}
\begin{tabular}{|c|c|c|c|c|c|c|c|c|c|}
 \noalign{\hrule height0.8pt}
 $R$ & $H_{(R)}$ & $ N_{\rm M}$ & $N_{\rm P}$ & NG & QNG &
 $H_{(R)}$-sector & {\footnotesize $\dim {\cal M}_R$} & $n_R$ & $N_R(G)$ \\
 \hline
 \noalign{\hrule height0.2pt}
  I & $SU(N-1)$ & $2N-1$ & $0$  & $2N-1$ & $2N-1$ & 
      $({\bf N-1})_{\rm M} \oplus {\bf 1}_{\rm M}$ & $1$ & $2$ &$2$ \\  
 II & $SU(N-2)$ & $2$   & $2N-3$ & $4N-4$ & $2$   & 
      $({\bf N-2})_{\rm P} \oplus {\bf 1}_{\rm P} \oplus 
       2 \,{\bf 1}_{\rm M}$ & $2$ & $2$ & $2$  \\
 \noalign{\hrule height0.8pt}
 \end{tabular}
 \end{center}
\end{table}
\end{footnotesize}

Since the generator $iX^{\rm diag.}\sim {\rm diag.}(1,\cdots,1,-(N-1))$ 
belongs to a singlet of $H_{\rm (I)}$, 
the dimension of region~I of the moduli space is 
$\dim{\cal M}_{\rm I}=1$ from theorem~2.
Since a transformation by $iX^{\rm diag.}$ 
mixes only broken generators, 
it does not change the structure of 
the $H$-sectors of the broken generators, 
and vacuum alignment does not occur.
It changes $\th_1$ and $\th_2$ while preserving $v\vt = f^2$; 
also, the orbit of this transformation in the moduli space 
is just a hyperbola, as shown in Fig.~8.
It is the boundary of the generic region (region~II).
The two generators $X_{N-1}^-$ and $X_{N-1}^+$ 
bring the vacuum in region~I to another region (region II).
Note that they do not commute with the generator $iX^{\rm diag.}$.

The K\"{a}hler potential contains 
two $G$-invariant variables in 
the arbitrary function as
\beq
 K = f({\th_1}^2,{\th_2}^2) 
   = f(\vec{\phi}\,\dagg  \vec{\phi} \;,\;
       \vec{\tilde{\phi}}\,\dagg \vec{\tilde{\phi}})
     |_{\vec{\tilde{\phi}}\cdot \vec{\phi} = f^2}
   = f(\vec{v}\,\dagg \xi\dagg \xi \vec{v} \;,\; 
       \vec{\vt}\,\dagg (\xi\dagg \xi)^{*-1} \vec{\vt}) \;,
\eeq
with a constraint $\vecvt \cdot \vec{v} = f^2$.

\subsection{Example of $\dim {\cal M} = 1$
with adjoint representation}

Until the last section, 
we have investigated only the fundamental representation.
This section and the next section are devoted to 
the adjoint representation.

\bigskip
Example 4) $SU(2)$ with $\vec{\phi} \in {\rm adj.}={\bf 3}$

First of all, we consider 
the simplest example $SU(2)$ with adjoint matter.
The fundamental fields are
\beq
 \phi = \phi^A T_A \;(A=1,2,3) \;\;,\; \phi^A \in {\bf C},
\eeq
where $T_A$ are related to the Pauli matrices as 
$T_A = \1{2} \sigma^A$.
They are traceless,
\beq
 {\rm tr}\phi = 0 \;. \label{trSU(2)}
\eeq
The transformation by $G^{\bf C}$ is 
\beq
 \phi \to  g \phi g^{-1} \;,\; 
 \phi\dagg \to  (g^{-1})\dagg \phi g\dagg \;,\;g \in G^{\bf C} \;.
   \label{gc}
\eeq

There is a matrix identity called the Cayley-Hamilton theorem,
\beq
 A^2 - (\tr A) A + (\det A) I_2 = 0 \;, \label{CH2}
\eeq
where $A$ is any two-by-two matrix and 
$I_2$ is a two-by-two unit matrix.
By putting $\phi$ in $A$, we obtain the identity, 
\beq
 \phi^2 =  - (\det\phi) I_2 \;. \label{tr-phi}
\eeq
~From Eq.~(\ref{trSU(2)}) and this equation, 
we can find that there is 
only one $G^{\bf C}$-invariant, $\tr \phi^2$.
By choosing a suitable superpotential, 
we obtain a F-term constraint,
\beq
 \tr \phi^2 - f^2 = 0 \;. \label{adj.SU(2)F}
\eeq 
Since $N(G^{\bf C}) = 1$, 
the complex dimension of the target space is 
$N_{\Ph} = \dim_{\bf C}V - N(G^{\bf C}) = 2$.
~From Eqs.~(\ref{tr-phi}) and (\ref{adj.SU(2)F}), 
we obtain 
\beq
 && \tr \phi^2 = - 2 \det\phi = f^2 \;,\label{formula1}\\ 
 && \phi^2 =  - (\det\phi) I_2 = \1{2} f^2 I_2 \;. \label{formula2}
\eeq

To find the independent $G$-invariant, 
we put $\phi\dagg\phi$ in $A$ in Eq.~(\ref{CH2}):
\beq   
  (\phi\dagg\phi)^2 
  &=& \tr(\phi\dagg\phi) \phi\dagg\phi 
      - \det(\phi\dagg\phi) I_2  \non
  &=& \tr(\phi\dagg\phi) \phi\dagg\phi - \1{4} f^4 I_2 \;,\label{formula3}
\eeq
where we used Eq.~(\ref{formula1}).
We can find that there is only one $G$-singlet: $N(G)=1$, 
\beq
 \tr (\phi\dagg \phi) \in {\bf R}\;,
\eeq
since from Eqs.~(\ref{formula2}) and (\ref{formula3}), 
other $G$-invariants,  
\beq
 && \tr (\phi\dagg \phi^2) = \1{2} f^2 \tr \phi\dagg = 0 \;,\\
 && \tr (\phi^{\dagger 2} \phi^2 )= \1{2} f^4 \;,\\
 && \tr((\phi\dagg\phi)^2) = (\tr(\phi\dagg\phi))^2 - \1{2} f^4  
\eeq
are not independent.

By using the $G^{\bf C}$ transformation, 
any vacuum can be transformed to the symmetric point in region~I,
\beq
 && v = <\phi> = {f \over \sqrt 2} \sigma_3 
      = {f \over \sqrt 2} \pmatrix{1 & 0 \cr 0 & -1} \;,\\
 && \tr v^2 = f^2 \;.
\eeq
The unbroken symmetry is $H_{\rm (I)}= U(1)$ and 
the broken generators are 
${\cal G}-{\cal H}_{\rm (I)}=\{\sigma_1,\sigma_2\}$.
The maximal realization occurs at the symmetric point: 
$G^{\bf C}/\hat H = SL(2,{\bf C})/GL(1,{\bf C})$ 
and $N_{\rm M} = 2\,,\,N_{\rm P} = 0$.
${\cal G}^{\bf C}-\hat{\cal H}$ belongs to 
a single $H_{\rm (I)}$-representation ${\bf 2}_{\rm M}$. 
Since $n_{\rm I}=1$ agrees with $N_{\rm I}(G)=1$,
the conjecture can be verified in region~I.
Since there is no singlet, 
the dimension of region~I is $\dim {\cal M}_{\rm I}=0$ from theorem~2.

The non-symmetric (generic) points belonging to region~II 
are written in 
\beq
 v \goto v\pri 
 = g_0 v {g_0}^{-1} 
 = a \sigma_3 + b \sigma_+ + c \sigma_-
 = \pmatrix{a & b \cr c & -a} \;, 
 a \in {\bf R} \;,\; b,c \in {\bf C}.
\eeq
The parameter $a$ is real, 
since any Hermitian matrix can be diagonalized 
by some unitary matrix $g \in G$.\footnote{
If we put $v= X+iY$, where $X$ and $Y$ are Hermitian matrices, 
$\tr v^2 = \tr (X^2 - Y^2 ) + 2 i \tr (XY) = f^2$.
By unitary transformation, 
$X$ can be diagonalized: $X \simeq \sigma_3$ 
and $Y \simeq \{\sigma_1,\sigma_2\}$.}
There is a relation, 
$\tr v^{\prime 2} = - 2 \det v\pri = 2 (a^2 + bc) = f^2$.
The real unbroken symmetry is trivial $\{1\}$, 
whereas there is complex unbroken symmetry 
$\pmatrix{a & b \cr c & -a} \in \hat{\cal H}$.
Thus in the complex broken generators 
there is one pure-type generator and 
the other is the mixed-type generator:
$N_{\rm M} = 1 \;,\; N_{\rm P} = 1$. 
Since both of them belong to singlet representations of $H_{\rm (II)}$, 
the dimension of region~II is $\dim {\cal M}_{\rm (II)}=1$ from theorem~2.
Since region II is the generic region, 
theorem~1 is $n_{\rm II}= N_{\rm II}(G)=1$.
\begin{table}
\caption{\bf phases of $SU(2)$ with ${\bf 3}$}
\begin{center}
\begin{tabular}{|c|c|c|c|c|c|c|c|c|c|}
 \noalign{\hrule height0.8pt}
 $R$ & $H_{(R)}$ & $ N_{\rm M}$ & $N_{\rm P}$ & NG & QNG &
 $H_{(R)}$-sector & {\footnotesize $\dim {\cal M}_R$} & $n_R$ & $N_R(G)$ \\
 \hline
 \noalign{\hrule height0.2pt}
  I & $U(1)$ & $2$ & $0$ & $2$ & $2$ & ${\bf 2}_{\rm M}$ & $0$ & $1$ & $1$\\  
 II & $\{1\}$  & $1$ & $1$ & $3$ & $1$ & 
      ${\bf 1}_{\rm P} \oplus {\bf 1}_{\rm M}$ & $1$ & $1$ & $1$  \\
 \noalign{\hrule height0.8pt}
 \end{tabular}
 \end{center}
\end{table}

There exist two phases, 
the symmetric region and non-symmetric region.
The moduli space is equivalent to that of Example~1, 
since there exists isomorphism 
$SU(2)/U(1) \simeq O(3)/O(2) \simeq S^2$.
Their complexifications are also equivalent. 

The effective K\"{a}hler potential can be written as
\beq
 K = f\left(\tr (\phi\dagg \phi) \right)|_{\tr \phi^2 = f^2} 
   = f\left(
            \tr ((\xi\dagg\xi)^{-1} v\dagg (\xi\dagg \xi) v) 
      \right) \;,
\eeq
with the constraint $\tr v^2 = f^2$, 
where we have used the fact  
$\phi|_{\rm F} = \xi v \xi^{-1}$ and its conjugate.

\subsection{Example of $\dim {\cal M} = 4$ 
with adjoint representation}
Example 5) $SU(3)$ with $\vec{\phi} \in {\rm adj.}={\bf 8}$

We now consider the $SU(3)$ case. 
The fundamental fields are 
\beq
 \phi = \phi^A T_A \;(A=1,\cdots,8) \;\;,\; \phi^A \in {\bf C},
\eeq
where $T_A$ are related to the Gell-Mann matrices as 
$T_A = \1{2} \lambda^A$.
They are traceless: ${\rm tr}\phi = 0$.
The transformation by $G^{\bf C}$ is same as Eq~(\ref{gc}).

The Cayley-Hamilton theorem is
\beq
 A^3 - {3 \over 2}(\tr A) A^2 + \1{2}(\tr A)^2 A + (\det A) I_3 = 0 \;,
 \label{CH3}
\eeq
where $A$ is any three-by-three matrix and 
$I_3$ is a three-by-three unit matrix.
By substituting $A = \phi$ in Eq.~(\ref{CH3}), 
we obtain
\beq
 \phi^3 =  - (\det\phi) I_3 \;. \label{phi3}
\eeq
~From this equation, 
we can find that there are 
two $G^{\bf C}$-invariants, $\tr \phi^2$ and $\tr \phi^3 = - 3\,\det\phi$, 
($N(G^{\bf C}) = 2$).
We can obtain F-term constraints from a suitable superpotential as 
\beq
 && \tr \phi^2 = \1{2} \delta_{AB} \phi^A \phi^B = f^2 \;,\\
 && \tr \phi^3 = - 3 \,\det\phi = d_{ABC} \phi^A \phi^B \phi^C =  g^3 \;,
 \label{GC-inv.}
\eeq
where $f$ and $g$ are real parameters and 
$d_{ABC} = \tr (\{\lambda_A, \lambda_B\}\lambda_C)$.
The complex dimension of the target space is 
$N_{\Ph} = \dim_{\bf C}V - N(G^{\bf C}) = 8-2 = 6$.

To investigate $G$-invariants, 
we put $A= \phi\dagg \phi$ in Eq.~(\ref{CH3}),
\beq
 (\phi\dagg\phi)^3 
&=& {3\over 2} \tr(\phi\dagg\phi) (\phi\dagg\phi)^2
 - \1{2} (\tr \phi\dagg\phi)^2 \phi\dagg\phi 
 - \det (\phi\dagg\phi) I_3 \non
&=& {3\over 2} \tr(\phi\dagg\phi) (\phi\dagg\phi)^2
 - \1{2} (\tr \phi\dagg\phi)^2 \phi\dagg\phi 
 + \1{9} g^6 I_3 \;, \label{SU3-formula}
\eeq
where we have used Eq.~(\ref{GC-inv.}).

~From Eqs.~(\ref{phi3}), (\ref{GC-inv.}) 
and (\ref{SU3-formula}), we find that 
there are four independent $G$-invariants: $N(G) = 4$,
\beq
 && \tr (\phi\dagg \phi) = \1{2}\delta_{AB} \phi^{*A}\phi^B \;,\; 
    \tr (\phi\dagg\phi)^2 \in {\bf R} \;,\non
 && \tr (\phi\dagg\phi^2) = d_{ABC} \phi^{*A}\phi^B\phi^C \in {\bf C}\;,
 \label{G-inv.SU(3)ad.}
\eeq
since the other $G$-invariants
\beq
 && \tr (\phi^{\dagger 2} \phi^2) 
 = \tr (\phi\dagg\phi)^2 - {3 \over 2}(\tr \phi\dagg\phi)^2 
   + {3 \over 2} \tr(\phi\dagg\phi) \;,\\
 && \tr (\phi\dagg \phi^3 ) = - (\det \phi) \tr \phi\dagg = 0 \;,\\
 && \tr(\phi^{\dagger 2} \phi^3) 
    = - (\det \phi) \,\tr\phi^{\dagger 2} = \1{3} f^2 g^3\\
 && \tr((\phi\dagg \phi)^2 \phi) 
    = {3 \over 2} \tr(\phi\dagg \phi) \tr (\phi\dagg\phi^2),\\
 && \tr(\phi\dagg \phi^4) 
    = -(\det\phi)\, \tr(\phi\dagg \phi) 
    =  \1{3} g^3 \tr(\phi\dagg \phi),\\
 && \tr ((\phi\dagg \phi)^3)
    = {3\over 2}\tr (\phi\dagg \phi) \tr((\phi\dagg\phi)^2) 
       -\1{2} (\tr (\phi\dagg \phi))^2 \tr (\phi\dagg \phi)
       + \1{3} g^6 ,\\
 && \cdots  \nonumber
\eeq
are all not independent.

The generic vacua can be transformed 
by a $G^{\bf C}$ transformation  
to the symmetric point, 
\beq
 v = <\phi> 
 = a \lambda_3 + b \lambda_8 
 = \pmatrix{a +{b \over \sqrt 3}   & 0 & 0 \cr
            0 & - a +{b \over \sqrt 3} & 0 \cr
            0 & 0    & -{2 \over \sqrt 3} b}  \;,\;
 a , b \in {\bf R}. \label{vsSU(3)}
\eeq
Since there are relations between parameters $a$ and $b$ 
and the value of the $G^{\bf C}$-invariants, $f$ and $g$ as
\beq
 \tr v^2 = 2(a^2 + b^2) = f^2 \;,\;
 \tr v^3 = 2\sqrt 3 \left(a^2 b - \1{3} b^3 \right) = g^3 \;,
\eeq
$a$ and $b$ are constant at the single $G^{\bf C}$-orbit.
Thus they also parametrize the $G^{\bf C}$-orbit space $V/G^{\bf C}$. 
There are five types of $G^{\bf C}$-orbits.
We list them in Table~5: (i)-(v). 
\begin{small}
\begin{table}
\caption{\bf $G^{\bf C}$-orbits of $SU(3)$ with ${\bf 8}$}
\begin{center}
\begin{tabular}{|c|c|c|c|c|c|}
 \noalign{\hrule height0.8pt}
 orbit & area & $v$ & ${\cal H}_{({\rm s})}$ &
 $H_{\rm (s)}$ & $N_{\Phi}$ \\
 \hline
 \noalign{\hrule height0.2pt}
  (i) & $a=b=0$ & $v = 0$ & 
   $\{\lambda_1,\cdots,\lambda_8\}$ & $SU(3)$ & $0$ \\ 
 (ii) & $a=0\;,b\neq 0$ & ${\rm diag.} 
   ({b \over \sqrt 3},{b \over \sqrt 3},-{2 \over \sqrt 3} b)$ & 
   $\{\lambda_1,\lambda_2,\lambda_3,\lambda_8\}$ &
   $SU(2)\times U(1)$ & $4$ \\  
(iii) & $a=\sqrt 3 b\;,\; b\neq 0$ & ${\rm diag.} 
   ({4 \over \sqrt 3}b, -{2 \over \sqrt 3}b, -{2 \over \sqrt 3} b)$ &
   $\{\lambda_3,\lambda_6,\lambda_7,\lambda_8\}$ & 
   $SU(2)\times U(1)$ & $4$\\  
 (iv) & $a=-\sqrt 3 b\;,\; b\neq 0$ & ${\rm diag.}
   (-{2 \over \sqrt 3}b, {4 \over \sqrt 3}b, -{2 \over \sqrt 3} b)$ & 
   $\{\lambda_3,\lambda_4,\lambda_5,\lambda_8\}$ & 
   $SU(2)\times U(1)$ & $4$ \\
(v) & generic & Eq.~(\ref{vsSU(3)}) & 
   $\{\lambda_3 , \lambda_8 \}$ & $U(1)^2$ & $6$ \\
 \noalign{\hrule height0.8pt}
 \end{tabular}
 \end{center}
\end{table}
\end{small}
In all cases, the maximal realization occurs, 
since the adjoint representation is a real representation. 
The last type contains the generic $G^{\bf C}$-orbits  
which have the maximal dimension 
and the others are singular $G^{\bf C}$-orbits 
with fewer dimensions than generic orbits.
In this paper 
we consider the generic $G^{\bf C}$-orbit.
Thus, the unbroken symmetry at the symmetric point (region I) is 
$H_{\rm (I)}= U(1)^2$ 
and the generators of it, 
${\cal H}_{\rm (I)} = \{\lambda_3 , \lambda_8 \}$ 
are the Cartan generators.
Since it is the maximal realization, 
the target space is 
$G^{\bf C}/\hat H = SL(3,{\bf C})/GL(1,{\bf C})^2$.
~From the commutation relations,
\beq
 &&[\lambda_3, \lambda_1] \sim \lambda_2 \;,\; 
   [\lambda_3, \lambda_2] \sim \lambda_1 \;,\;
   [\lambda_8, \lambda_1] \sim 0 \;,\; 
   [\lambda_8, \lambda_2] \sim 0 \;, \non
 &&[\lambda_3, \lambda_4] \sim \lambda_5 \;,\; 
   [\lambda_3, \lambda_5] \sim \lambda_4 \;,\;
   [\lambda_8, \lambda_4] \sim \lambda_5 \;,\; 
   [\lambda_8, \lambda_5] \sim \lambda_4 \;,\non
 &&[\lambda_3, \lambda_6] \sim \lambda_7 \;,\; 
   [\lambda_3, \lambda_7] \sim \lambda_6 \;,\;
   [\lambda_8, \lambda_6] \sim \lambda_7 \;,\; 
   [\lambda_8, \lambda_7] \sim \lambda_6 \;,\label{lambda-com.}
\eeq
we can find 
the $H_{\rm (I)}$-representations 
to which the complex broken generators belong, as follows.
$\lambda_1 \oplus \lambda_2$ belongs to 
$({\bf 2}_{\rm M},{\bf 1}_{\rm M})$ from the first line of 
Eq.~(\ref{lambda-com.}) and 
$\lambda_4 \oplus \lambda_5$ and $\lambda_6 \oplus \lambda_7$
belong to $({\bf 2}_{\rm M},{\bf 2}_{\rm M})$ from 
the second and third lines of Eq.~(\ref{lambda-com.}).
Here, $(\cdot,\cdot)$ denotes the representation of 
the unbroken symmetry generated by $(\lambda_3,\lambda_8)$.
There are three sectors: $n_{\rm I} = 3$, 
which coincides with the number of the $G$-invariants, 
since at this point $\phi$ becomes Hermitian, $\phi\dagg = \phi$, 
so the third type of the $G$-invariants 
in Eq.(\ref{G-inv.SU(3)ad.}) becomes real 
and $N_{\rm I}(G) = 3$. 
Thus the conjecture is nontrivially also true 
in this region of the model.
Since there is no $H_{\rm (I)}$-singlet, 
the dimension of region~I is $\dim {\cal M}_{\rm (I)} = 0$ 
from theorem~2.

We transform the symmetric vacuum $v_{\rm (I)}$ 
to another vacuum $v_{\rm (II)}$ as  
\beq
 v_{\rm (I)} \goto v_{\rm (II)} 
 = g_0 v_{(I)}{g_0}^{-1} \;,\; 
 g_0 = \exp( \alpha E_{(+,0)} + \beta E_{(-,0)})  \in  G^{\bf C}\;,
\eeq
where $\alpha$ and $\beta$ are some real parameters and 
\beq
 E_{(\pm,0)} 
 = (T_1 \pm i T_2) 
 = \1{2}\pmatrix{0 & 1 & 0 \cr
                 0 & 0 & 0 \cr
                 0 & 0 & 0 } \;,\;
   \1{2}\pmatrix{0 & 0 & 0 \cr
                 1 & 0 & 0 \cr
                 0 & 0 & 0 } 
\eeq
are broken generators in the Cartan form, 
whose root vectors are $(\pm,0)$. 
Since the unbroken generators are transformed to 
\beq
&& \lambda_3 \goto g_0 \lambda_3 {g_0}^{-1} 
 = a\pri \lambda_3 + c \lambda_+ + d \lambda_-  \;,\; \label{trlam3}
   a\pri \in {\bf R}\;,\; c,d\in {\bf C}\\
&&  \lambda_8 \goto g_0 \lambda_8 {g_0}^{-1} = \lambda_8 ,\label{trlam8}
\eeq
the obtained vacuum is in the form 
\beq
 v_{\rm (II)} 
 = a \lambda_3 + b \lambda_8 + c \lambda_+ + d \lambda_-
 = \pmatrix{a\pri +{b \over \sqrt 3}   & c & 0 \cr
            d & - a\pri +{b \over \sqrt 3} & 0 \cr
            0 & 0      & -{2 \over \sqrt 3} b} \;. \label{SU(3)v2}
\eeq
Note that $cd \in {\bf R}$, since 
$\det \phi = 
 (-{2 \over \sqrt 3})\{({b \over \sqrt 3})^2 - {a\pri}^2 - cd\}
 = - \1{3} g^3 \in {\bf R}$.
Therefore, from a short calculation, 
the third type of $G$-invariant in Eq.~(\ref{G-inv.SU(3)ad.}), 
which is complex at general point, 
also remains real in region~II, 
since $\tr(\phi\dagg\phi^2) 
= 2(-a\pri+ {b \over \sqrt 3})^3 + (-{2 \over \sqrt 3} b)^3 
 + {2 \over \sqrt 3}b(|c|^2 + |d|^2 + cd) \in {\bf R}$.
Thus, the number of $G$-invariants is $N_{\rm II}(G)=3$.

The unbroken symmetry in this region is $H_{\rm (II)} = U(1)$, 
generated by $\lambda_8$.
~From Eqs.~(\ref{trlam3}) and (\ref{trlam8}), 
$\lambda_1$ is transformed to a non-Hermitian generator and 
$\lambda_8$ remains Hermitian.
Since there is one more complex unbroken generator 
in addition to the real unbroken generators, 
there is one pure-type complex broken generator 
besides five mixed-type broken generators.
The $H_{\rm (II)}$-sectors of the complex broken generators 
are as follows:
$(\lambda_4 ,\lambda_5)$ and $(\lambda_6 ,\lambda_7)$ 
belong to ${\bf 2}_{\rm M}$, 
one of the complex combination of 
$\lambda_1,\lambda_2$ and $\lambda_3$ to ${\bf 1}_{\rm M}$
and another combination to ${\bf 1}_{\rm P}$.
(The rest combination is in ${\hat H}_{\rm (II)}$ 
as noted above.)
Thus, the number of mixed $H_{\rm (II)}$-sectors is $n_{\rm II}=3$, 
whivh is in agreement with $N_{\rm II}(G)$. 
Therefore, the conjecture is also nontrivially verified in region~II.
The results are given in the second line of Table~6.
\begin{table}
\caption{\bf phases of $SU(3)$ with ${\bf 8}$}
\begin{center}
\begin{tabular}{|c|c|c|c|c|c|c|c|c|c|}
 \noalign{\hrule height0.8pt}
 $R$ & $H_{(R)}$ & $ N_{\rm M}$ & $N_{\rm P}$ & NG & QNG &
 $H_{(R)}$-sector & {\footnotesize $\dim {\cal M}_R$} & $n_R$ & $N_R(G)$ \\
 \hline
 \noalign{\hrule height0.2pt}
   I & $U(1)^2$ & $6$ & $0$ & $6$ & $6$ & 
       $({\bf 2},{\bf 1})_{\rm M} \oplus 2({\bf 2},{\bf 2})_{\rm M}$ & 
       $0$ & $3$ & $3$  \\  
  II & $U(1)$ & $5$ & $1$ & $7$ & $5$ & 
       ${\bf 1}_{\rm P} \oplus 2\, {\bf 2}_{\rm M}
       \oplus {\bf 1}_{\rm M}$ & $1$ & $3$ & $3$  \\
 III & $\{1\}$  & $4$ & $2$ & $8$ & $4$ & 
       $2\, {\bf 1}_{\rm P} \oplus 4 \,{\bf 1}_{\rm M}$ & $4$ & $4$ & $4$  \\
 \noalign{\hrule height0.8pt}
 \end{tabular}
 \end{center}
\end{table}

There exists one more region as a generic region (region III).
The most generic vacuum can be written as
\beq
 v_{\rm (III)} = a \lambda_3 + b \lambda_8 + \sum_{A \neq 3,8}  c_A \lambda_A 
 \;\;,\; a,b \in {\bf R} \;,\; c_A \in {\bf C} \;.
\eeq
Although this vacuum breaks the symmetry $G$ completely, 
there exist two complex unbroken generators corresponding 
to $\lambda_3$ and $\lambda_8$ in the symmetric point.
Thus the complex broken generators are constituted from 
two pure-type and six mixed-type generators, 
and all of them belong to singlets, 
since there is no unbroken symmetry,  
as is the third line of Table~6.

The effective K\"{a}hler potential can be written as
\beq
 K 
 &=& f\left( \tr(\phi\dagg \phi)\,,\, \tr(\phi\dagg\phi)^2 \,,\,
        \tr(\phi\dagg \phi^2 + \phi\phi^{\dagger 2} ) \,,\,
        -i\tr(\phi\dagg \phi^2 - \phi\phi^{\dagger 2} )
      \right) |_{\tr \phi^2 = f^2\,,\,\tr \phi^3 = g^3} \non
 &=& f[ 
      \tr\left((\xi\dagg\xi)^{-1} v\dagg 
               (\xi\dagg\xi) v \right) \,,\, 
      \tr\left((\xi\dagg\xi)^{-1} v\dagg 
               (\xi\dagg\xi) v \right)^2 \,,\non
 &&\hspace{0.4cm}  
      \tr\left((\xi\dagg\xi)^{-1} v\dagg (\xi\dagg\xi) v^2 
       + (\xi\dagg\xi) v (\xi\dagg\xi)^{-1} v^{\dagger 2}\right)
      \,,\non
&&\hspace{0.3cm}  
    -i \tr\left((\xi\dagg\xi)^{-1} v\dagg (\xi\dagg\xi) v^2 
        - (\xi\dagg\xi) v 
          (\xi\dagg\xi)^{-1} v^{\dagger 2}\right) 
       ] \;,
\eeq
with constraints $\tr v^2 = f^2\,,\,\tr v^3 = g^3$.

In this example, 
since it is nontrivially verified 
that $N_R(G)$ changes accordingly to $n_R$, 
we believe that the conjecture is generically true.

\section{Comments on gauging global symmetry}
In this paper, we have mainly considered 
a theory which has only global symmetry. 
We comment in this section 
on a theory with gauge symmetry. 
If a theory has a gauge symmetry, 
we consider it as being global symmetry for a while, 
and gauge it after finding the moduli space.
If the theory has global and gauge symmetry, 
we require partial gauging of the global symmetry, 
while if it has only gauge symmetry, 
we require full gauging of the global symmetry.

First of all, we consider the case that 
all of the global symmetry $G$ is gauged.
The gauging brings the D-flat condition, 
$(\vec{\phi}\dagg T_A \vec{\phi})^2 = 0$, 
besides the F-flat condition. 
This condition can be replaced by the condition that 
the length $|\vec{\phi}|$ is minimum~\cite{D-flat,GS}.
It chooses the one $G$-orbit (if it exists) from 
the set of F-flat points, namely the target space $M$.
(It is called the D-orbit in Ref.~\cite{LT}.)
It is known that a closed $G^{\bf C}$-orbit has one 
D-flat $G$-orbit~\cite{D-flat,GS,LT}. 
For example, in Example~1 
the $G^{\bf C}$-orbit is closed and 
there exists one D-orbit where the moduli parameter 
$\th_1 = |\vec{\phi}|_{\rm F}$ is minimum. 
(See Fig.~3 and Fig.~4.) 
They are a set of symmetric points. 
It has been proved that 
the complex unbroken symmetry $\hat H$ is reductive, 
namely $\hat H = H^{\bf C}$, and that there is no Borel algebra 
in the D-flat orbit~\cite{D-flat,GS}. 
Actually, the symmetric points have this property and 
the maximal realization occurs there. 
In moduli space, the D-orbit corresponds to one point. 
Thus, moduli space is trivial. 
In Example~1, it is the point labeled D in Fig.~6. 
In example~3, the symmetric region, where $\hat H$ is reductive, 
has one dimension. 
The D-point is one of them, 
where the length of the vector 
$\vec{\Phi} \defeq (\vec{\phi},\vec{\tilde{\phi}})$ 
in a reducible representation of $G$ is minimum. 
The minimum of 
$|\vec{\Phi}|^2 = |\vec{\phi}|^2 + |\vec{\tilde{\phi}}|^2 
= (\th_1)^2 + (\th_2)^2$ 
is shown in Fig.~8.
The D-orbit is the fibre at D-point in the moduli space.

However, if the $G^{\bf C}$-orbit is an open set, 
there is no D-flat point. 
In Example~2, the $G^{\bf C}$-orbit is open and 
there is no D-flat point. 
Since there exist Borel subalgebra, 
there is no point such that $\hat H$ becomes reductive. 
Thus supersymmetry must be broken spontaneously.

It is known that supersymmetry is spontaneously broken 
in a gauged sigma model 
with only pure-type multiplets~\cite{gauging,WB,Ku}. 
Only when the maximal realization can occur, 
supersymmtery is preserved. 
There is a physical explanation to this phenomenon~\cite{Ku}.
When the massless vector superfields absorb 
the NG chiral superfields, 
if there exist pure-type multiplets, 
they cannot constitute massive vector multiplets 
and supersymmetry must be spontaneously broken. 

In our method, 
it is sufficient 
to see whether the $G^{\bf C}$-orbit is open or closed, 
instead investigating whether supersymmetry is broken or not.
Thus, our method may be useful to investigate 
dynamical supersymmetry breaking. 
(For a review, see Ref.~\cite{DSB}.)

\bigskip
The case of partial gauging is more complicated. 
In the general embedding case, 
the ordinary vacuum alignment problem occurs 
besides the supersymmetric vacuum alignment~\cite{KS2}.   
However, when the gauged group is an ideal,  
namely the whole symmetry is the direct product of 
the global symmetry and the gauge symmetry,~\footnote{
We assume the whole symmetry is compact group.} 
it does not occur. 
Since this includes the case of 
the supersymmetric QCD~\cite{KS,Fe}, 
it is interesting to consider this case. 
We leave it to future works.

\section{Conclusions}

The moduli space of the gauge symmetry in 
$N=1$ supersymmetric theory is well understood. 
However, for the case of global symmetry, 
it was not known at all.
We have investigated 
the moduli space of the $N=1$ supersymmetric theory 
with only global symmetry. 
In the case of global symmetry, 
although the complexified group is a symmetry of 
the superpotential, and thus the F-term scalar potential, 
since it contains only chiral superfields, 
it is not the symmetry of the D-term potential, 
since the D-term contains chiral and anti-chiral superfields. 
Thus, the moduli space of global symmetry 
is the quotient space of the set of F-flat points 
divided by the symmetry $G$.

On the other hand, 
it has been known that 
when the global symmetry $G$ spontaneously 
brakes down to $H$ while preserving $N=1$ supersymmetry, 
the low-energy effective Lagrangian is 
the nonlinear sigma model 
with the K\"{a}hler target manifold $M=G^{\bf C}/\hat H$ 
parametrized by NG and QNG bosons. 
The target manifold is embedded in 
the space of the fundamental fields. 
Since the target manifold $M \simeq G^{\bf C}/\hat H$ 
just comprises the F-flat points, 
the moduli space is ${\cal M}= (G^{\bf C}/\hat H) /G$.
An investigation in this direction requires  
a deep understanding of the K\"{a}hler coset manifolds 
which had not yet been done.

It has been known that 
there is a supersymmetric vacuum alignment 
in this type of theory.
The target manifold has non-compact directions 
corresponding to the appearance of the QNG bosons. 
The vacuum degeneracy in this non-compact direction 
has a one-to-one correspondence with 
the freedom to embed 
the complex unbroken symmetry $\hat H_v$ to $G^{\bf C}$. 
Therefore, the unbroken symmetry 
$H_v = \hat H_v \cap G$ depends on 
the points $\vec{v}$ in the target manifold.
(We have called the symmetric point $\vec{v}$ 
the point with the largest real unbroken group $H_v$ and 
the generic point the point with the least symmetry. )
The number of NG and QNG bosons changes from point to point 
with the total number of NG and QNG bosons being unchanged.
The compact coset manifolds $G/H_v$ ($G$-orbit of $\vec{v}$), 
with various dimensions, parametrized by the NG bosons 
are embedded in the full target manifold $M$.

The K\"{a}hler potential of the 
low-energy effective Lagrangian 
which describes the low energy behavior of 
the NG and QNG bosons 
can be written as 
the arbitrary function of 
some moduli parameters. 
By identification of the fundamental fields with 
the F-term constraints and the K\"{a}hler coset representative, 
we have found that it coincides with 
the known K\"{a}hler potential 
constructed by a group-theoretical way.

\bigskip
We have decomposed the moduli space into some regions ${\cal M}_R$,
such that the $real$ unbroken symmetries 
at different points in the same region 
are isomorphic to each other by a $G^{\bf C}$-transformation.
We have investigated the moduli space by 
differential geometric (or the group theoretical) view points, 
such as the K\"{a}hler coset manifold, 
and by algebraic geometric view points such as 
the ring of $G$-invariant polynomials.

\medskip
~From the differential geometrical view points,  
the moduli space is obtained by 
the identification of $G$-orbits in 
the full target manifold.
Thus, the target manifold $M$ is considered to be 
a fibre bundle with fiber $G$-orbits $G/H_v$ on 
the base moduli space ${\cal M}$. 
At the boundary region $R_2$ of some region $R_1$ ($\del R_1 = R_2$), 
the unbroken symmetry $H_{(R_1)}$ is enhanced to $H_{(R_2)}$. 
This corresponds to that $H_{(R_1)}$-orbit in $G$-orbit shrinks and 
most of the NG bosons change to the QNG bosons there. 
(The NG-QNG change occurs.)
The symmetric points of target space correspond to 
the most singular point of the moduli space, 
and the number of QNG bosons is maximal.
On the other hand, at generic points of moduli space, 
the space of QNG bosons is identical to the tangent vector space 
of the moduli space, $T_p{\cal M}$, and 
the number of QNG bosons is minimum with agreement with
the dimension of the moduli space.

The complex broken generators can be decomposed to 
$H_{(R)}$-irreducible sectors, 
since they are transformed linearly 
by the action of $H_{(R)}$.
The number of the $H_{(R)}$-irreducible sectors is
the number of independent non-compact directions. 
They also correspond to the directions of moduli space. 
It can change at each region. 

\medskip
~From algebraic geometrical view points, 
the ring of $G^{\bf C}$-invariant polynomials 
is generated by the finite $G^{\bf C}$-invariants and 
the target manifold is obtained 
by fixing all of them. 
We have considered the generic $G^{\bf C}$-orbits. 
On the other hand, 
the ring of the $G$-invariant polynomials is 
also generated by the finite $G$-invariant polynomial 
and the $G$-orbit is obtained by fixing all of them. 
So the moduli space ${\cal M}$ is 
parametrized by such the $G$-invariants 
(after fixing the $G^{\bf C}$-invariants). 

~From the relation of these two methods, 
we have obtained theorem~1 (Eq.~(\ref{theorem1})) 
which states that, in the generic region, 
the number of the $G$-invariants coincides with 
the number of the $H_{(R)}$-irreducible sectors of mixed types.
We have also conjectured that 
it is true in any region, 
since both quantities are equal to 
the number of independent non-compact directions. 
We indeed show that this is true in examples, especially in example 5. 
We have also obtained theorem~2 (Eq.~(\ref{theorem2})) 
concerning the dimension of the regions of the moduli space 
denoted that $\dim {\cal M}_R$ coincides with 
the number of $H_{(R)}$-singlet sectors of mixed types. 
~From these theorems, we have obtained 
formulae (\ref{corollary2}) to calculate 
the dimension of the moduli space in various ways. 

\bigskip
We have examined the results in many examples 
using the method of the algebraic geometry and 
the differential geometry (or the group theory).

When the fields belong to the fundamental representation, 
it is quite easy to calculate the dimension of 
the moduli space by using the former method.
However it is difficult to calculate it by the latter method, 
since we must classify the complex broken generators 
to the pure- and mixed-types and 
study their transformation properties under unbroken symmetry. 

On the other hand, 
when the fields belong to the adjoint representation, 
it is quite easy to calculate the dimension of 
the moduli space by using the latter method, 
since the unbroken symmetry is just the Cartan subalgebra.
However, it is quite difficult to calculate it by the former method, 
since we must use freely 
the Cayley-Hamilton theorem to 
reduce the number of $G$-invariants. 

\bigskip
In this paper, we have considered only 
the generic $G^{\bf C}$-orbits.
Generalization to 
the other $G^{\bf C}$-orbits is straightforward.
By considering it, 
it is possible to generalize to vacua 
with non-transitive $G^{\bf C}$ action. 
(In such theories, there are extra flat directions 
not with related to the symmetry breaking.) 

We hope that our method is useful 
to construct the supersymmetric Wess-Zumino term~\cite{SWZ}, 
to satisfy anomaly matching, 
to consider dynamical supersymmetry breaking~\cite{DSB}, 
to investigate the effective action of 
the branes in curved space~\cite{BC} 
and other theoretical and phenomenological subjects 
of the modern physics.
We hope to return to these subjects in future studies.

\section*{Acknowledgements}
We thank K.~Higashijima for 
useful discussions, encouragement and 
carefull reading of the manuscript.  
We are grateful to K.~Ohta and N.~Ohta for 
arguments during the early stage of this work.
We also thank T.~Yokono for some useful comments  
on the moduli space of the supersymmetric gauge theories 
and M.~Goto for some comments on the orbit space.



\begin{thebibliography}{99}

\bibitem{HKLR}
N.~J.~Hitchin, A.~Karlhede, U.~Lindst\"{o}m and M.~Ro\v{c}ek, 
Comm. Math. Phys. {\bf 108} (1987) 535.

\bibitem{LT}
M.~A.~Luty and W.~Taylor IV, 
Phys. Rev. {\bf D53} (1996) 3339, hep-th/9506098.

\bibitem{DM}
G.~D.~Dotti and A.~V.~Manohar, 
{\em Anomaly Matching Conditions and the Moduli Space of 
Supersymmetric Gauge Theories}, hep-th/9710024. 

\bibitem{Zu}
B.~Zumino,
Phys. Lett. {\bf 87B} (1979) 203.

\bibitem{BKMU}
M.~Bando, T.~Kuramoto, T.~Maskawa and S.~Uehara,
Phys. Lett. {\bf 138B} (1984) 94;
Prog. Theor. Phys. {\bf 72} (1984) 313, 1207.

\bibitem{Le}
W.~Lerche,
Nucl. Phys. {\bf B238} (1984) 582.

\bibitem{BL}
W.~Buchm\"{u}ller and W.~Lerche, 
Ann. Phys. {\bf 175} (1987) 159.

\bibitem{Sh}
G.~M.~Shore,
Nucl. Phys. {\bf B320} (1989) 202;  
Nucl. Phys. {\bf B334} (1990) 172.  

\bibitem{KS}
A.~C.~Kotcheff and G.~M.~Shore,
Int. J. Mod. Phys. {\bf A4} (1989) 4391;
Nucl. Phys. {\bf B333} (1990) 701;  
Nucl. Phys. {\bf B336} (1990) 245.  

\bibitem{LRM}
M.~A.~Luty, J.~March-Russel and H.~Murayama,  
Phys. Rev. {\bf D52} (1995) 1190, hep-ph/9501233.

\bibitem{IKK}
K.~Itoh, T.~Kugo and H.~Kunitomo,
Nucl. Phys. {\bf B263} (1986) 295;
Prog. Theor. Phys. {\bf 75} (1986) 386.

\bibitem{BE}
W.~Buchm\"{u}ller and U.~Ellwanger, 
Phys. Lett. {\bf 166B} (1985) 325.

\bibitem{HNOO}
K.~Higashijima, M.~Nitta, K.~Ohta and N.~Ohta, 
Prog. Theor. Phys. {\bf 98} (1997) 1165, hep-th/9706219.

\bibitem{HN}
K.~Higashijima and M.~Nitta, 
{\em Geometry of $N=1$ Supersymmetric Low Energy Theorems}, 
KEK-TH~571, to appear.

\bibitem{Ku}
T.~Kugo, 
Soryuusiron Kenkyuu (Kyoto) {\bf 95} (1997) C56.

\bibitem{BKY}
M.~Bando, T.~Kugo and K.~Yamawaki,
Phys. Rep. {\bf 164} (1988) 217.

\bibitem{KS2}
A.~C.~Kotcheff and G.~M.~Shore, 
Nucl. Phys. {\bf B301} (1988) 267.  

\bibitem{D-flat}
R.~Gatto and G.~Sartori, 
Phys. Lett. {\bf 118B} (1982) 79; 
G.~Girardi, P.~Sorba and R.~Stora, 
Phys. Lett. {\bf 144B} (1984) 212; 
C.~Procesi and G.~W.~Schwarz, 
Phys. Lett. {\bf 161B} (1985) 117; 
R.~Gatto and G.~Sartori, 
Phys. Lett. {\bf 157B} (1985) 389. 

\bibitem{GS}
R.~Gatto and G.~Sartori, 
Comm. Math. Phys. {\bf 109} (1987) 327.

\bibitem{AS}
M.~Abud and G.~Sartori, 
Phys. Lett. {\bf 104B} (1981) 147; 
Ann. Phys. {\bf 150} (1983) 307.

\bibitem{gauging}
J.~Bagger and E.~Witten, 
Phys. Lett. {\bf 118B} (1982) 103;  
A.~J.~Buras and W.~Slominski, 
Nucl. Phys. {\bf B223} (1983) 157;  
C.~M.~Hull, A.~Karlhede, U.~Lindstr\"{o}m and M.~Ro\v{c}ek, 
Nucl. Phys. {\bf B266} (1986) 1;  
J.~Bagger and J.~Wess, 
Phys. Lett. {\bf 199B} (1987) 243; 
E.~J.~Chun, 
Phys. Rev. {\bf D41} (1990) 2003.

\bibitem{SWZ}
D.~Nemeschansky and R.~Rohm, 
Nucl. Phys. {\bf B249} (1985) 157; 
E.~Cohen and C.~G\'{o}mez,  
Nucl. Phys. {\bf B254} (1985) 235; 
S.~Aoyama and J.~W.~van Holten,  
Nucl. Phys. {\bf B258} (1985) 18. 


\bibitem{Fe}
F.~Feruglio, {\sl A smooth massless limit for supersymmetric QCD}, 
hep-th/9802178.


\bibitem{DSB}
E.~Poppitz and S.~P.~Trivedi, {\sl Dynamical Supersymmetry Breaking}, 
hep-th/9803107

\bibitem{BC}
M.~R.~Douglas, {\em D-branes in Curved Space}, hep-th/9703056; 
M.~R.~Douglas, A.~Kato and H.~Ooguri, 
{\em D-brane Actions on K\"{a}hler manifolds}, hep-th/9708012.

\bibitem{WB}
J.~Wess and J.~Bagger,
{\em Supersymmetry and Supergravity}, Princeton Univ. Press, Princeton(1992).

\end{thebibliography}
\end{document}